\documentclass[12pt]{article}
\usepackage{epsfig}
\usepackage{url}
\newcommand{\ba}{\begin{array}}
\newcommand{\ea}{\end{array}}
\newcommand{\bd}{\begin{displaymath}}
\newcommand{\ed}{\end{displaymath}}
\newcommand{\be}{\begin{equation}}
\newcommand{\ee}{\end{equation}}
\newcommand{\bea}{\begin{eqnarray}}
\newcommand{\eea}{\end{eqnarray}}
\newcommand{\Dir}{\kern -6.4pt\Big{/}}
\newcommand{\Dirin}{\kern -10.4pt\Big{/}\kern 4.4pt}
\newcommand{\DDir}{\kern -10.6pt\Big{/}}
\newcommand{\DGir}{\kern -6.0pt\Big{/}}
\begin{document}
\
\def\bra{\langle}
\def\ket{\rangle}

\def\a{\alpha}
\def\as {\alpha_s}
\def\b{\beta}
\def\d{\delta}
\def\e{\epsilon}
\def\ve{\varepsilon}
\def\l{\lambda}
\def\m{\mu}
\def\n{\nu}
\def\G{\Gamma}
\def\D{\Delta}
\def\L{\Lambda}
\def\s{\sigma}
\def\p{\pi}

\def\etal{ {\em et al.}}
\def\mzs {M_Z^2}
\def\mws {M_W^2}
\def\q2 {q^2}
\def\sz {\sin^2\theta_W}
\def\cz {\cos^2\theta_W}
\def\lp{\lambda^{\prime}}
\def\lps{\lambda^{\prime *}}
\def\lpp{\lambda^{\prime\prime}}
\def\lpps{\lambda^{\prime\prime * }}

\def\bapp{b_1^{\prime\prime}}
\def\bbpp{b_2^{\prime\prime}}
\def\bcp{b_3^{\prime}}
\def\bdp{b_4^{\prime}}
\def\t {\times }
\def\slash {\!\!\!\!\!\!/}
\def\photino {\tilde\gamma}
\def\sel {\tilde{e}}
 \def\N10{\widetilde \chi_1^0}
                         \def\C1p{\widetilde \chi_1^+}
                         \def\C1m{\widetilde \chi_1^-}
                         \def\C1pm{\widetilde \chi_1^\pm}
 \def\Ntwo{\widetilde \chi_2^0}
                         \def\Ctwo{\widetilde \chi_2^\pm}
\def\lslep {{\tilde e}_L}
\def\rslep {{\tilde e}_R}
\def\sneu {\tilde \nu}
\def\msneu {M_\tilde \nu}
\def\mrslep {m_{\rslep}}
\def\mlslep {m_{\lslep}}
\def\mneu {m_{\neu}}
\def\mpT{p_T \hspace{-1em}/\;\:}
\def\mET{E_T \hspace{-1.1em}/\;\:}
\def\mE{E \hspace{-.7em}/\;\:}
\def\go{\rightarrow}
\def\beq{\begin{eqnarray}}
\def\Rp{R\!\!\!\!/}
\def\wrp {{\cal W}_{R\!\!\!\!/}}
\def\enq{\end{eqnarray}}
\def\goes{\longrightarrow}
\def\lsim{\:\raisebox{-0.5ex}{$\stackrel{\textstyle<}{\sim}$}\:}
\def\gsim{\:\raisebox{-0.5ex}{$\stackrel{\textstyle>}{\sim}$}\:}

 \begin{flushright}
{SHEP-07-11}\\
{FNT/T 2007-04}\\
{\today}\\
\end{flushright}
\begin{center}
{\large\bf Vector-Boson Production of\\[2mm]
 Light Higgs Pairs in 2-Higgs Doublet Models}\\[3mm]
{\large M. Moretti}\\
{\em  Dipartimento di Fisica, Universit\`a di Ferrara and\\
INFN - Sezione di Ferrara, Via Paradiso 12, 44100 Ferrara, Italy
}\\[2mm]
{\large S. Moretti}\\
{\em School of Physics \& Astronomy, University of Southampton,\\
Highfield, Southampton SO17 1BJ, UK, and}\\
{\em Laboratoire de Physique Th\'eorique, Universit\'e Paris--Sud,
F--91405 Orsay Cedex, France}
\\[2mm]
{\large F. Piccinini}\\
{\em INFN - Sezione di Pavia, Dipartimento di Fisica Nucleare 
   e Teorica,\\
    Via Bassi 6, 27100 Pavia,
Italy}
\\[2mm]
{\large R. Pittau\footnote{Present address:
{\em Institute of Nuclear Physics, NCSR "DEMOKRITOS",
    15310, Athens, Greece.}}}\\
{\em  Dipartimento di Fisica Teorica, Universit\`a di Torino and \\
INFN 
  - Sezione di Torino,  Via Giuria 1, 10125 Torino, Italy, and}\\
{\em Departamento de F\'{i}sica Te\'orica y del Cosmos,\\
 Centro Andaluz de F\'{i}sica de Part\'{i}culas Elementales (CAFPE),\\
 Universidad de Granada, E-18071 Granada, Spain}
 \\[2mm]
{\large J. Rathsman}\\
{\em High Energy Physics, Uppsala University,
Box 535, 751 21 Uppsala, Sweden}
\\[1mm]
\end{center}
\begin{abstract}
{\noindent 
At the Large Hadron Collider, we prove the 
feasibility to detect 
pair production of 
the lightest CP-even Higgs boson $h$ of
Type II 2-Higgs Doublet Models through
$q q^{(')}\to q q^{(')} {hh}$ (vector-boson fusion).
We also show that, through the $hh\to 4b$ decay
channel in presence of heavy-flavour tagging, further exploiting 
forward/backward jet sampling,
one has direct access to the 
$\lambda_{Hhh}$ triple Higgs coupling
-- which constrains the form of the Higgs potential.
}
  
\end{abstract}


\newpage

\section{Introduction}\vspace*{0.25cm}\noindent

\label{sec:intro}

If only a light Higgs boson (with mass $M_{h}\lsim140$ GeV) is found at the
Large Hadron Collider (LHC), it may be difficult to tell whether
it belongs to the Standard Model (SM) or indeed a model with an
enlarged Higgs sector. For example, in the case of a CP-conserving
Type II 2-Higgs Doublet Model (2HDM) 
\cite{guide}--\cite{Abdel}\footnote{Of the
initial eight degrees of freedom pertaining to the two complex Higgs
doublets, only five survive as real particles upon Electro-Weak 
Symmetry Breaking (EWSB), labelled as $h, H$, $A$
(the first two are CP-even or `scalars' (with $M_h<M_H$) whereas the third is CP-odd or
`pseudoscalar') and $H^\pm$, as three degrees of freedom 
are absorbed into the definition
of the longitudinal polarisation for the gauge bosons $Z$ and $W^\pm$,
upon their mass generation after EWSB.}, possibly in presence of minimal 
Supersymmetry (SUSY) -- the 
combination of the two yielding the so-called Minimal 
Supersymmetric Standard Model (MSSM) -- this happens in the so-called
`decoupling region', when $M_{H},M_{A},M_{H^\pm}\gg M_{h}$, 
for suitable
choices of the other MSSM and 2HDM parameters, where - for the same mass - the $h$ couplings
to ordinary matter in the SM are the same as in both the 2HDM and MSSM. 
Even in these conditions, however, it has been proved that one could
possibly establish the presence of an extended Higgs sector by determining
the size of the trilinear Higgs self-coupling $\lambda_{hhh}$ \cite{HHH}.

If the extended model is not in a decoupling condition, then it is generally
possible to establish the presence of additional Higgs signals,
$H,{A}$ and/or ${H^\pm}$ \cite{ATLAS,CMS}. However, even when this
is the case, it may be difficult to distinguish, e.g., between a
generic Type II 2HDM and the MSSM (unless, of course, one also detects the SUSY
partners of ordinary matter and Higgs bosons). In fact, despite there
exist well establish spectra among the four different masses in the MSSM
(for fixed, say, $M_{h}$ and $\tan\beta$, the ratio of the vacuum expectation
values of the two Higgs doublets in either model),
it may well be possible that the additional 2HDM parameters arrange
themselves to produce an identical mass pattern. However, such a degeneracy
between the two models would not typically
persist if one were able to also measure certain Higgs couplings, chiefly 
those among the Higgs bosons themselves (involving two or more
such particles). In fact, while the measurement
of only two among the four Higgs boson masses
($M_{h},M_{H},M_{A}$ and $M_{H^\pm}$) -- or, alternatively,
one such masses and $\tan\beta$ -- would fix (at tree-level)
all Higgs masses and couplings in the MSSM, this is no longer true in a generic
Type II 2HDM \cite{guide}, because of the freedom in selecting the
free additional parameters. For example, the general CP-conserving Type II 2HDM 
that we are going to consider can be specified uniquely by seven
parameters: 
$M_{h},M_{H},M_{A}$, $M_{H^\pm}$, $\beta$, $\alpha$ (the mixing angle between
the two CP-even neutral Higgs states) and
$\lambda_5$ (see  
eq.~(\ref{eq:potential}) later on). It may then 
happen that the first
six of these are measured and found to agree with the MSSM pattern, but one
would still need to measure $\lambda_5$ to verify that it is the Higgs
sector of the MSSM that is present. One way to do so would be by 
measuring trilinear Higgs self-couplings, such as $\lambda_{hhh}$ 
and $\lambda_{Hhh}$. Alternatively, the
measurement of the latter two couplings would constitute a test of the MSSM
relations if one knew $M_{h}$ and $\tan\beta$ but not $\alpha$.

In this paper, we make the assumption that only one parameter is known,
$M_{h}$, as may well happen at the LHC 
after only a $h$ resonance is detected. We further imply that all (potential) SUSY
states are much heavier than ordinary particles (with the possible exception
of the lightest SUSY particle, see footnote 4), thus effectively a decoupled
MSSM setup. Under these circumstances, we then ask ourselves the following
question. While trying to establish the presence of additional (single) heavy
Higgs signals, which would then unmistakably distinguish between the
SM and a scenario with an extended Higgs sector, would it also be possible
to gather information on Higgs self-couplings from signatures involving two 
light Higgs bosons,
hence by studying channels involving $h$ pair production, thereby
possibly also distinguish between, e.g., a generic Type II 2HDM and
the MSSM? 

\begin{figure}[!t]
  ~\hskip1.0cm\epsfig{file=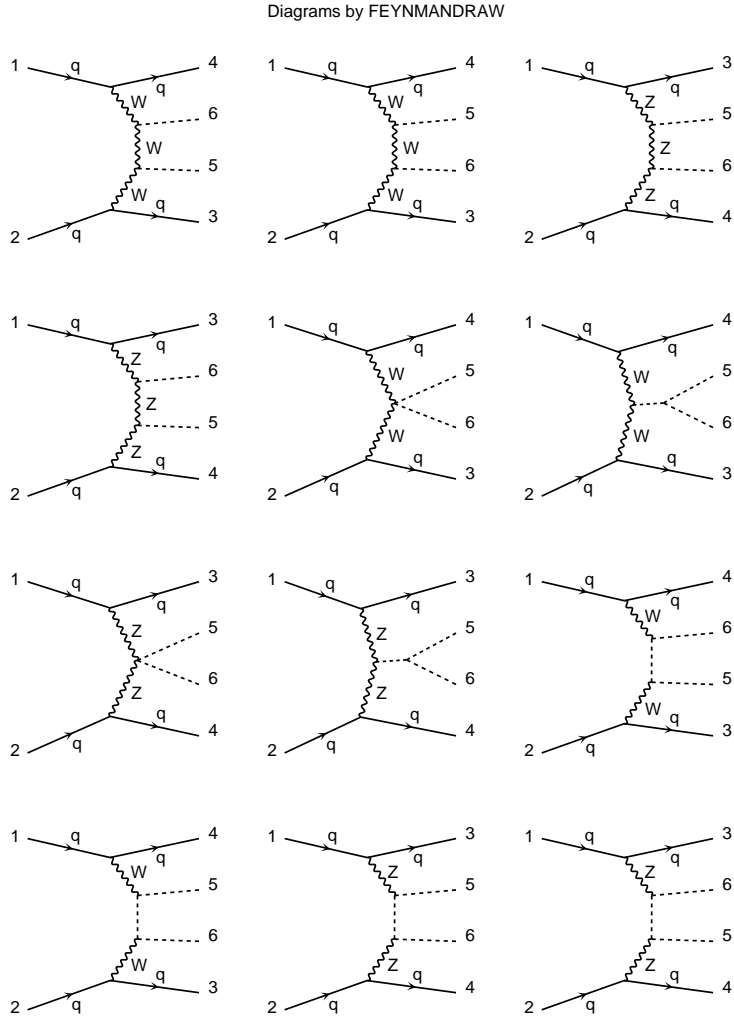,height=18cm,angle=0}
\vspace*{-3cm}
\caption{Feynman diagrams for $q_1q_2\to q_3q_4 h_5h_6$.
Depending on the (anti)quark flavour combination, the $W^\pm$- and
$Z$-mediated graphs may not interfere. Besides, for final state
(anti)quarks  of different flavours, only half of the diagrams survive.}
\label{fig:VVgraphs}
\end{figure}

It is the purpose of this paper to show that this is the case, so long
that enough luminosity can be accumulated at the LHC, also
in view of the Super-LHC (SLHC) option \cite{SLHC}. We will illustrate 
how we have come to this conclusion, i.e., after investigating the process
\cite{VVHH}
\bea\label{procs}
q q^{(')}&\to& q q^{(')} {hh} ~ ({\mathrm{vector-boson~fusion}}),
\eea
with $q^{(')}$ referring to any possible (anti)quark
flavour combinations\footnote{The gluon-gluon production mode 
\cite{ggHH} was considered
in Refs.~\cite{Remi} and \cite{LesHouches1999}
(see also \cite{standard}), and later on \cite{Baur1,Remi2},
where -- despite significant kinematic differences exist between signal and QCD noise --
it was eventually shown that the extraction of the $gg\to hh\to b\bar b 
b\bar b$ signal is essentially impossible at the (S)LHC because of the overwhelming QCD noise,
both reducible and irreducible.
Recently, encouraging results on the cross-section for multi-Higgs boson 
production in the gluon-gluon production mode has been obtained in models 
beyond the SM and MSSM~\cite{Binoth:2006ym}. 
The possibility of using Higgs boson pair
production more generally to access trilinear Higgs couplings has also been 
studied on the level of total cross-sections in~\cite{Djouadi:1999rca}.}. 
The relevant Feynman diagrams corresponding to process (\ref{procs}) in both
the MSSM and 2HDM considered here can be 
found in Fig.~\ref{fig:VVgraphs}. In our selection analysis, we 
will resort to the extraction of two
$h\to b\bar b$ resonances, in presence of  the following signature:
\begin{itemize}
\item `four $b$-quark jets and two forward/backward-jets'.
\end{itemize}
This signature was already considered in Ref.~\cite{HHH}
in the SM context (from which we will import some of the results). 

Our paper is organised as follows. In the next section, we outline
the computational procedure. Sect.~3 presents our numerical results
and discusses these in various subsections. Sect.~4 contains our conclusions.

\section{Calculation}
\label{sec:calcul}

We have assumed $\sqrt s=14$ TeV for the LHC energy throughout. 
Our numerical results are obtained by setting the renormalisation and 
factorisation scales to $2M_{h}$
for the signal while for the QCD
background we have used the average jet transverse momentum 
($p_T^2=\sum_{1}^{n}p_{Tj}^2/n$).
Both Higgs processes and noise were 
estimated by using the Parton Distribution Function (PDF) set MRST99(COR01)
\cite{hepdata}. 
While the background 
calculations were based on exact tree-level Matrix
Elements (MEs) using the ALPGEN program \cite{ALPGEN},
all signal rates were obtained through the same level of accuracy via
 programs
based on the HELAS subroutines
\cite{HELAS} -- for the computation of the MEs -- 
and VEGAS \cite{VEGAS} or Metropolis
\cite{Metropolis} -- for the multi-dimensional
integrations over the phase space. 
As for numerical input values of SM parameters,
we adopted the ALPGEN defaults. 

Concerning the MSSM setup, the two independent tree-level parameters that
we adopt are $M_A$ and $\tan\beta$. Through higher orders, 
we have considered the so called `Maximal
Mixing' scenario ($X_t=A_t-\mu/\tan\beta =\sqrt6 M_{\rm{SUSY}}$) \cite{MaxMix}, 
wherein 
we have chosen for the relevant SUSY input parameters: 
$\mu=200$ GeV, $A_b=0$, with 
$M_{\rm{SUSY}}=5$ TeV, the latter -- as already intimated --
implying a sufficiently heavy scale for all sparticle masses, so
that these are not accessible at the LHC and
no significant interplay between the SUSY and Higgs sectors of the model
can take place\footnote{The only 
possible exception in this mass hierarchy would be the
Lightest Supersymmetric Particle (LSP), whose mass may well be smaller
than the lightest Higgs mass values that we will be considering. However,
we have verified that invisible $h$ decays (including the one
into two LSPs) have negligible decay rates.}. Masses and couplings
within the MSSM have been obtained by using the {HDECAY} program \cite{HDECAY}.

Before giving the details of the 2HDM setup we are using, let us recall the
most general CP-conserving 2HDM scalar potential which is symmetric under 
$\Phi_{1(2)}\to-\Phi_{1(2)}$ up to softly breaking
dimension-2 terms (thereby allowing for loop-induced flavour 
changing neutral currents)~\cite{guide},
\begin{eqnarray}\label{eq:potential}
V &=& 
m_{11}^2\Phi_1^\dagger\Phi_1
+
m_{22}^2\Phi_2^\dagger\Phi_2
-
\left\{m_{12}^2\Phi_1^\dagger\Phi_2 + h.c.\right\}
+
\frac{1}{2}\lambda_1\left(\Phi_1^\dagger\Phi_1\right)^2
+
\frac{1}{2}\lambda_2\left(\Phi_2^\dagger\Phi_2\right)^2
+
\nonumber \\  &&
+
\lambda_3\left(\Phi_1^\dagger\Phi_1\right)\left(\Phi_2^\dagger\Phi_2\right)
+
\lambda_4\left(\Phi_1^\dagger\Phi_2\right)\left(\Phi_2^\dagger\Phi_1\right)
+
\left\{ \frac{1}{2}\lambda_5 \left(\Phi_1^\dagger\Phi_2\right)^2  + h.c.\right\}.
\end{eqnarray}
In the following,
the parameters $m_{11}$, $m_{22}$, $m_{12}$, $\lambda_1$, $\lambda_2$, 
$\lambda_3$ and $\lambda_4$ are replaced by
$v$, $M_{h}$, $M_{H}$, $M_{A}$, $M_{H^\pm}$, $\beta$ and $\alpha$ 
(with $v$ fixed). 
Hence, as intimated already, the CP-conserving 2HDM potential is parameterised by 
seven free parameters. 
Notice that 
from the scalar potential all the different Higgs couplings needed for our study
can easily be obtained. (See~\cite{THDM,Fawzi} for a 
complete compilation of couplings in a general CP-conserving 2HDM.)

In our 2HDM, we will
fix $M_{h}$ and $M_{H}$ to values similar to the ones found in the
MSSM scenario we are considering, by adopting
three different setups:
\begin{enumerate}
\item
$M_h=115$ GeV, $M_{H}=300$ GeV,
\item
$M_h=115$ GeV, $M_{H}=500$ GeV,
\item
$M_h=115$ GeV, $M_{H}=700$ GeV.
\end{enumerate}
We always scan over the remaining parameters
in the ranges
\begin{eqnarray*}
&-\pi/2<\alpha<\pi/2,&\\
&-4\pi<\lambda_5<4\pi,&\\
&0<\tan\beta<50,&\\
&100 \mbox{ GeV} < M_{A} <1000 \mbox{ GeV}, &\\
&100 \mbox{ GeV} <  M_{H^\pm}<1000 \mbox{ GeV}. &
\end{eqnarray*}

In order to accept a point from the scan we also check 
that the following conditions are fulfilled: the potential is bounded
from below, 
the $\lambda_i$ fulfill the tree-level unitarity 
constraints of \cite{Akeroyd:2000wc} and yield a contribution 
to $|\Delta\rho| < 10^{-3}$.
In short the unitarity constraints amounts to putting limits on the eigen 
values of the $S$ matrices for scattering various combinations of Higgs and
electroweak gauge bosons. We have followed the normal 
procedure~\cite{guide} of requiring the $J=0$ partial waves ($a_0$) of 
the different scattering processes to fulfill  $|\rm{Re}(a_0)|<1/2$, which 
corresponds to applying the condition that the eigenvalues\footnote{Here, 
$Z_2$ refers to the $Z_2$ symmetry, $Y$ is the hypercharge, and 
$\vec{\sigma}$ is the total weak isospin} 
$\Lambda_{Y\sigma\pm}^{Z_2}$ of the scattering matrices (or more precisely 
$16 \pi S$) fulfill $|\Lambda_{Y\sigma\pm}^{Z_2}| < 8 \pi$~\cite{Ginzburg:2005dt}. 
In other words we allow parameter space points all the way up to
the tree-level unitarity constraint $|\rm{Re}(a_0)|<1/2$. In order to 
investigate the sensitivity to this upper limit we will also report results 
as a function of the value of the maximal eigenvalue, $\Lambda_{\max}$.
The spectrum of masses, couplings and decay rates
in our 2HDM is the same as in Ref.~\cite{charged_Higgs_pairs},
obtained by using a modification of {HDECAY} \cite{HDECAY}
(consistent with a similar manipulation of the program used in
Ref.~\cite{BRs}). 
For each accepted point in the scan the partial decay rates 
for the different Higgs bosons are then calculated using {HDECAY}
and also taking possible additional partial widths of the 
${H}$ into account.

While the parameter dependence of the MSSM Higgs sector renders
the computation of the tree-level MSSM cross-sections rather straightforward
(as the latter depends on two parameters only, $M_{A}$ and $\tan\beta$),
the task becomes much more time-consuming in the context of the 2HDM.
In order to calculate the cross-sections in this scenario, they are
schematically written as a combination of couplings and kinematic factors 
in the following way:
\begin{equation}
\sigma_{\rm tot} = \int \left| \sum_{i=1}^{5} g_i M_i 
\right|^2 d {\rm{LIPS}} = 
\sum_{i=1}^{5}\sum_{j=i}^{5} g_ig_j\sigma_{ij},
\end{equation}
where all the explicit 
dependence on $\alpha$, $\beta$, $\lambda_{Hhh}$ and 
$\lambda_{hhh}$ is contained in the couplings $g_i$:  
$ g_1= \sin^2(\beta-\alpha)$, 
$ g_2 = \cos^2(\beta-\alpha) $, 
$ g_3 = \cos(\beta-\alpha)\lambda_{Hhh}$, 
$ g_4 = \sin(\beta-\alpha)\lambda_{hhh}$, and 
$ g_5 = 1$,
whereas the 
dependence on masses and other couplings is in the factors 
\begin{equation}
\sigma_{ij}=\frac{1}{1+\delta_{ij}}\int \left( M_i^\dagger M_j + 
M_j^\dagger M_i \right) d {\rm{LIPS}}.
\end{equation}
Note that the sum over subamplitudes $M_i$ also contains all interference 
terms and that colour factors etc.\ are included properly.\footnote{We have
carefully verified the integrity of our procedure.}
The $\sigma_{ij}$ are then calculated numerically for fixed masses. 
We can then get the cross-section in an arbitrary parameter space point by
multiplying the kinematic factors with the appropriate couplings. However,
there is a slight complication since the kinematic factor for the
$H \to hh$ contribution depends on the width $\Gamma_{H}$ if 
there is a s-channel resonance and the width in turn depends on the couplings. 
In this case the kinematic factor scales as $1/\Gamma_{H}$ which is 
accounted for by assuming a fixed value for the width when the kinematic factor
is calculated and then rescaling the result with the true width when calculating
the contribution to the cross-section. 
Another complication is the dependence of the kinematic factors 
on the Higgs masses, $M_{A}$ and $M_{H^\pm}$. The contributions of main
interest, which contain the $\lambda_{Hhh}$ and 
$\lambda_{hhh}$ couplings, only depend on these masses indirectly through
the unitarity constraints. At the same time there are other 
contributions to the cross-section which depend explicitly on these masses. 
However, these contributions are
very small in the parts of parameter space of interest and can thus be safely 
neglected.

\section{Results}
\label{sec:results}

In our investigation of the emerging hadronic
final state, we will 
assume that $b$-quark jets are distinguishable from light-quark and gluon
ones and neglect considering $b$-jet charge determination. 
Finite calorimeter resolution
has been emulated through a Gaussian smearing in transverse momentum,
$p_T$, with $(\sigma(p_T)/p_T)^2=(0.60/\sqrt{p_T})^2 +(0.04)^2$, for all 
jets. The corresponding missing transverse momentum, 
$p_T^{\mathrm{miss}}$, was reconstructed from the vector sum of 
the visible momenta after resolution smearing. Finally,
in our parton level analysis, we 
have identified jets with the 
partons from which they originate and applied all cuts directly to the
latter, since parton shower and hadronisation effects were not included in our
study.

\subsection{Inclusive Signal Results}
\label{subsec:inclusive}

In this section, after a preliminary analysis of the Higgs mass and coupling spectra
in the MSSM and a general Type II 2HDM, we will start our numerical analysis by investigating the model 
parameter dependence of the Higgs pair production process in (\ref{procs})
at fully inclusive level, in presence of the decay of the latter into
two $b\bar b$ pairs, with the integration over the  phase space being performed
with no kinematical restrictions. This will be followed by an analysis of the
production and decay process pertaining to the Higgs signal of interest at fully differential
level, in presence of detector acceptance cuts and kinematical selection constraints.
Finally, we will compare the yield of the signal to that of the corresponding
background and perform a dedicated signal-to-background study including an
optimisation of the cuts in order to enhance the overall significance. 
We will treat the MSSM and 2HDM in two separate subsections.
 
\subsubsection{MSSM}
\label{subsubsec:MSSM}

\begin{figure}[!ht]
  ~\hskip0.0cm\epsfig{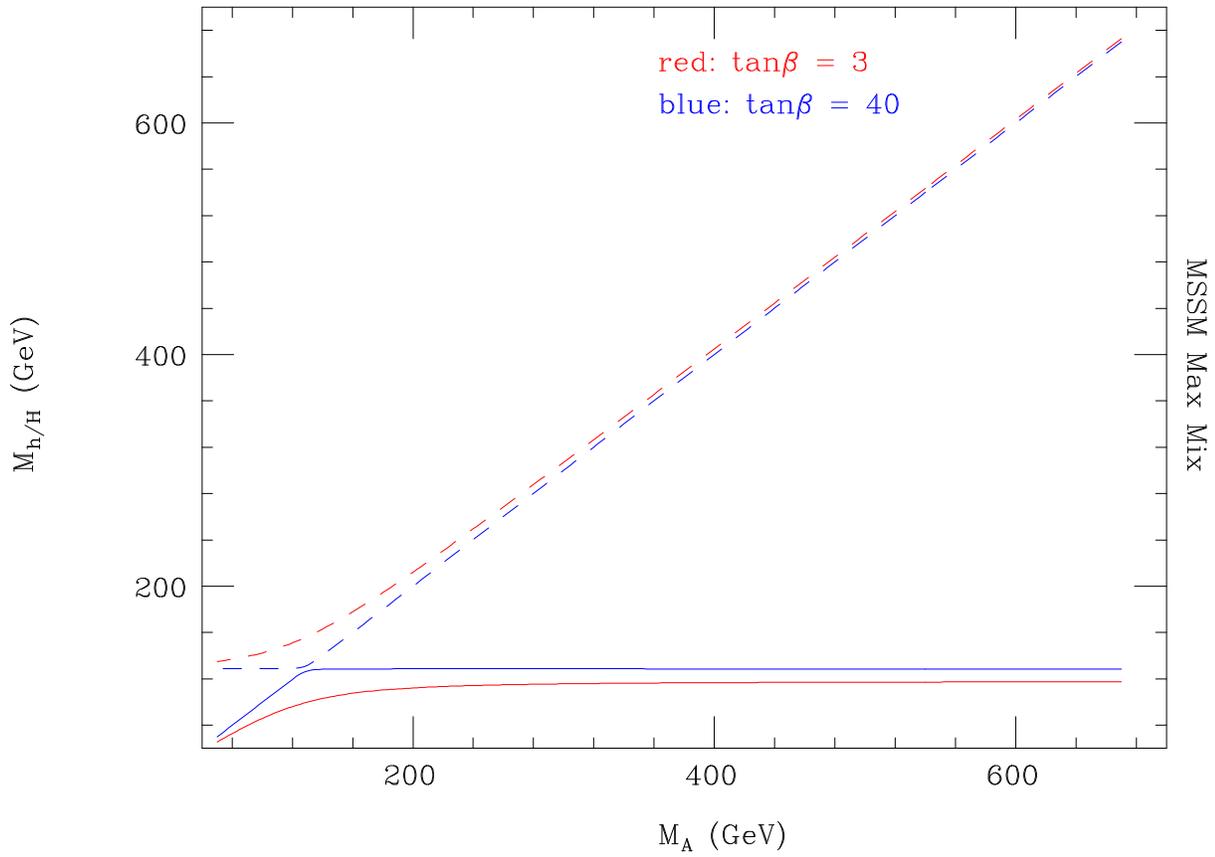}
\caption{The masses of the neutral CP-even Higgs bosons as a function of 
 the CP-odd one, for two choices of $\tan\beta$, assuming the MSSM in Maximal
Mixing configuration.}
\label{fig:massesMSSM_MaxMix}
\end{figure}

\begin{figure}[!ht]
  ~\hskip0.0cm\epsfig{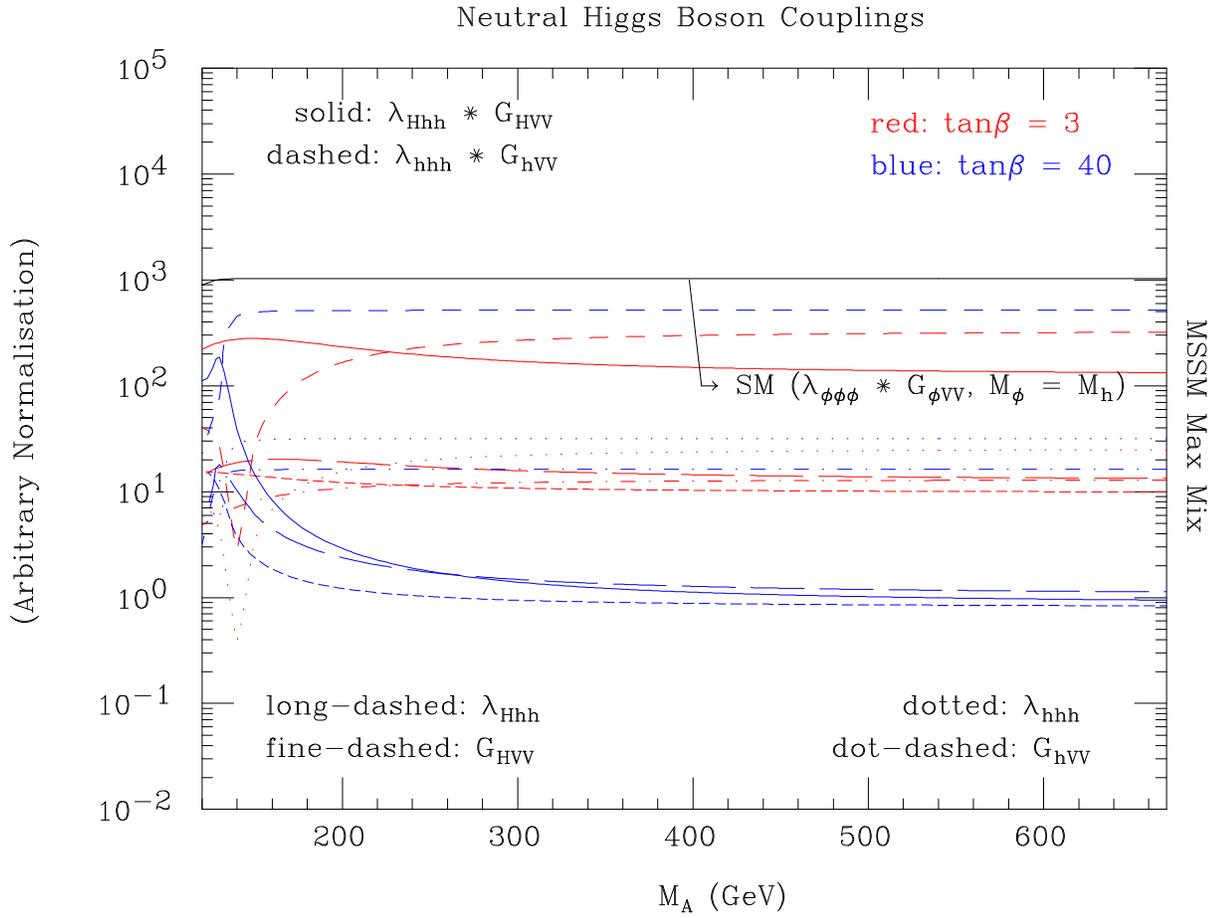}
\caption{The relevant couplings of the neutral CP-even Higgs bosons entering
the production process in (\ref{procs}) as a function of 
 the CP-odd Higgs boson  mass, for two choices of $\tan\beta$, assuming the MSSM in Maximal
Mixing configuration.}
\label{fig:couplingsMSSM_MaxMix}
\end{figure}

As representative of the low and high $\tan\beta$ regime, we will use in the remainder
the values of 3 and 40. We have instead treated $M_{A}$ as a continuous parameter,
varying between 100 and 700 GeV or so\footnote{Values of $M_{A}$ below 90 GeV or so
are actually excluded by LEP for the lower $\tan\beta$ value: see \cite{HiggsLEP2}.}.
Before proceeding with the numerical analysis of the signal, it is worthwhile to
investigate both the Higgs mass and coupling dependence in the MSSM with respect to the
two input parameters $M_{A}$ and $\tan\beta$. This is done in Figs.~\ref{fig:massesMSSM_MaxMix}
and \ref{fig:couplingsMSSM_MaxMix}, respectively. In the latter, 
we study  the case of (CP-even) MSSM Higgs boson couplings to gauge bosons
(denoted by $G_{hVV}$ and $G_{HVV}$), wherein $V$ refers to either a $W^\pm$
or a $Z$. In the same figure, the symbol $\phi$ refers to the SM Higgs boson, with
mass identical to that of the lightest MSSM Higgs state ($M_\phi=M_h$). 
While the pattern of masses has been well established in past literature, it is interesting to notice
here that the product of the MSSM couplings entering process (1) 
is always smaller than in the SM case.
However, in the MSSM, resonance enhancements can occur (such as in
$H\to hh$), so that the actual
MSSM production rates can in some cases overcome the corresponding SM
ones (for $M_\phi=M_h$). 

\begin{figure}[!ht]
  ~\hskip0.0cm\epsfig{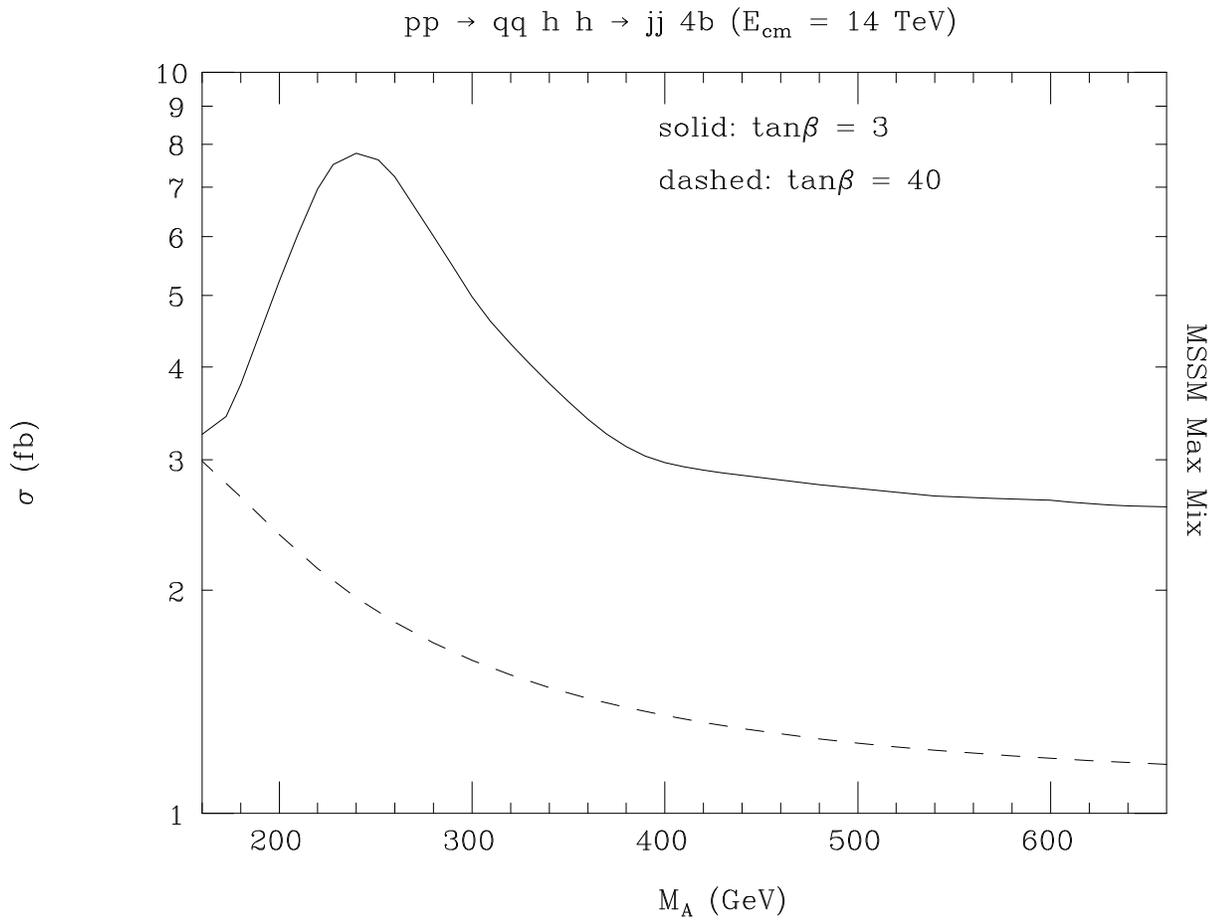}
\caption{The inclusive cross-sections  (as defined in the text) for
vector-boson fusion in (\ref{procs}), followed by
$hh\to b\bar bb\bar b$ decays, as a function of 
 the CP-odd Higgs boson  mass, for two choices of $\tan\beta$, 
assuming the MSSM in Maximal Mixing configuration.}
\label{fig:scan_qqhh_MSSM_MaxMix}
\end{figure}

Fig.~\ref{fig:scan_qqhh_MSSM_MaxMix}
presents the fully inclusive MSSM cross-section for the process
of interest, as defined in (\ref{procs}), times (effectively)
BR($hh\to b\bar b b\bar b$). The shape of the curves is mainly dictated
by the interplay between phase space (see Fig.~\ref{fig:massesMSSM_MaxMix})
and coupling (see Fig.~\ref{fig:couplingsMSSM_MaxMix}) effects, with the
exception of the region $M_A\gsim 220$  GeV and $\tan\beta=3$, where the
onset of the $H\to hh$ resonance is clearly visible. Cross-sections are generally sizable,
particularly at low $\tan\beta$. The displayed rates however coincide to the ideal situation
in which all final state jets  are 
detected with unit efficiency and the detector coverage extend to their
entire phase space, so that they only serve as a guidance in rating the phenomenological
relevance of the process discussed.

A more realistic analysis is in order, which we have performed as follows.
The four $b$-jets emerging from the decay of the $hh$ pair are accepted
according to the following criteria:
\begin{equation}\label{precuts1}
p_T^b>30~{\rm{GeV}},\qquad 
\vert \eta^b \vert < 2.5,\qquad
\Delta R_{bb} > 0.7,
\end{equation}
in transverse momentum, pseudorapidity and cone separation, respectively.
Their tagging efficiency is taken as $\epsilon_{b}=50\% $ for each $ b $ satisfying
these requirements, $ \epsilon_{b}=0 $ otherwise\footnote{Here and in the remainder,
the label $b$ refers to jets that are $b$-tagged while $j$ to any jet (even those
originating from $b$-quarks)
which is not.}. In addition,
to enforce the reconstruction of the two Higgs
bosons, we require all such $b$'s in the event to be tagged 
and that at least one out of the three possible double 
pairings of $b$-jets satisfies the following mass preselection:
\be 
\label{eq:mhchi2}
(m_{b_1,b_2}-M_h)^2+(m_{b_3,b_4}-M_h)^2 < 2~\sigma_m^2, 
\ee 
where $\sigma_m= 0.12~ M_h$. We further exploit
`forward/backward-jet' tagging, by imposing that the non-$b$-jets
satisfy the additional cuts
\begin{equation}\label{precuts2}
 p_T^{\rm{fwd/bwd}} >  20~{\rm GeV},\qquad
  2.5 <  \eta^{\rm{fwd}}  <  5,\qquad
 -2.5 >  \eta^{\rm{bwd}}  > -5.
\end{equation}

Tab.~\ref{tab:lambda} shows the rates of the signal 
after the implementation of the constraints in eqs.~(\ref{precuts1})--(\ref{precuts2})
(hereafter, referred to as `acceptance and preselection cuts' or `primary cuts'). While
our process does yield non-negligible rates after the latter, it turns out 
that it
is of no phenomenological relevance, even assuming
very high luminosity. Firstly, in view of the fact that $b$-tagging efficiencies
are not taken into account in this table: for the `$4b$-jet' tagging 
option, one should multiply the
numbers in Tab.~\ref{tab:lambda} by $\epsilon_b^4$, that is, 1/16. 
(Alternative approaches requiring a lesser number of $b$-jets to be tagged
as such were not successful either.)
Secondly,
the background rates, after the same cuts in eqs.~(\ref{precuts1})--(\ref{precuts2}),
are always overwhelming the signal, despite our efforts in further optimising the cuts.
For this reason, rather than dwelling upon the latter now, we postpone their discussion
to the next subsection and simply conclude here that
our channel is altogether inaccessible at both the LHC and SLHC
in the context of the MSSM. 

 \begin{table}[t]
\begin{center}
\begin{tabular}{|c|c|c|c|}
\hline
\multicolumn{4}{|c|}{$\tan\beta=3 $} \\
\hline
$M_A$ (GeV)  &   $M_h$ (GeV)  &  $\sigma(qq^{(')}\to qq^{(')}hh) $ [fb]    & $\sigma({\rm{background}}) $ [fb]  \\
\hline
160   &  108   & 0.19    &   218  \\
200   &  112   & 0.23    &   232  \\
240   &  114   & 0.46    &   229  \\
\hline
\hline
\multicolumn{4}{|c|}{$\tan\beta=40 $} \\
\hline
$M_A$ (GeV)  &   $M_h$ (GeV)  &  $\sigma(qq^{(')}\to qq^{(')}hh) $ [fb]    & $\sigma({\rm{background}}) $ [fb]  \\
\hline
160   &  129   & 0.26    &   224  \\
200   &  129   & 0.20    &   224  \\
240   &  129   & 0.17    &   224  \\
\hline
\end{tabular}
\end{center}
\caption{Cross-sections for Higgs pair production via vector-boson fusion, 
process (\ref{procs}), after Higgs boson decays (relevant BRs are all included)
and the acceptance and preselection cuts defined
in (\ref{precuts1})--(\ref{precuts2}), for two choices of $\tan\beta$ and 
a selection of $M_{A}$ values, assuming the MSSM
in Maximal Mixing configuration
(the corresponding values of $M_h$ are also 
indicated in brackets).  No $b$-tagging efficiencies are
included here. }
\label{tab:lambda}
\end{table}

\subsubsection{2HDM}
\label{subsec:2HDM}

\begin{figure}
\center
\epsfig{file=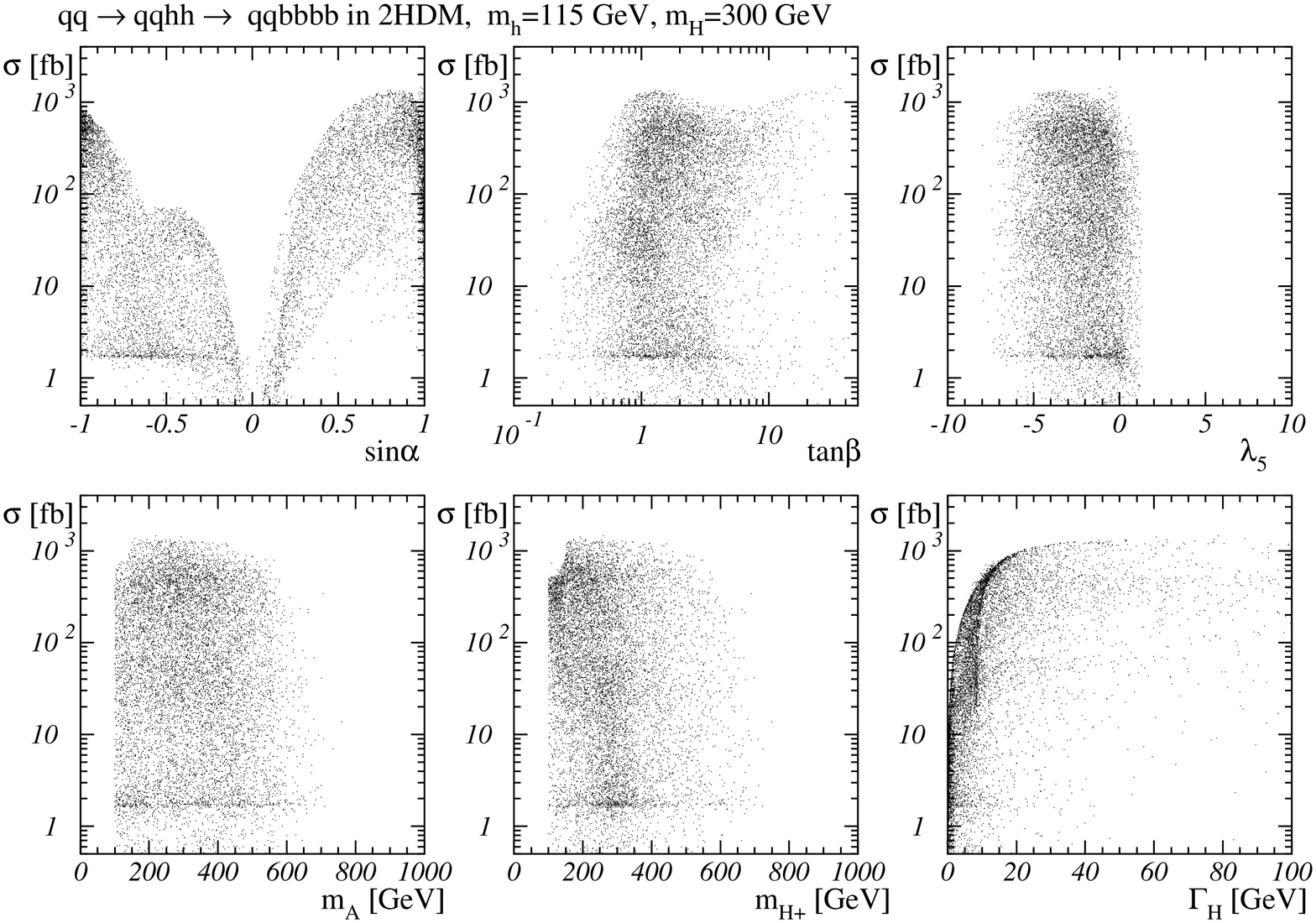,  height=12cm}
\caption{ The dependence of the  inclusive cross-section 
$qq^{(')}\to qq^{(')}hh \to
qq^{(')}b\bar bb\bar b$   in the 2HDM under consideration on the different parameters when
scanning over 10000 parameter space points 
 for $M_h=115$ GeV and $M_{H}=300$ GeV.  (Note that $M_A$ and $M_{H^\pm}$ are
free parameters.)}
\label{fig:dsig_dpar_nocut_300}
\end{figure}

\begin{figure}
\center
\epsfig{file=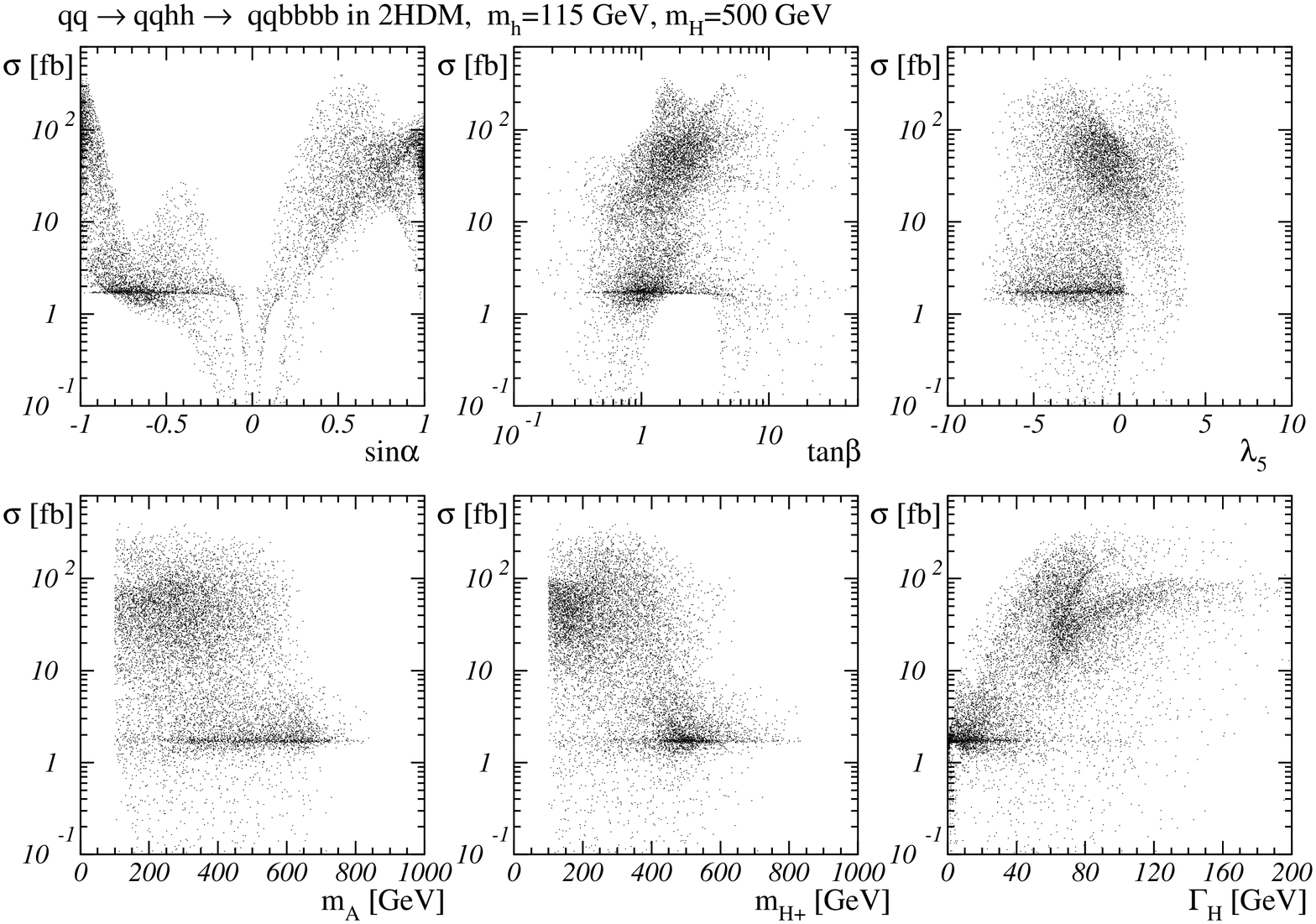,  height=12cm}
\caption{ The dependence of the  inclusive cross-section 
$qq^{(')}\to qq^{(')}hh \to
qq^{(')}b\bar bb\bar b$   in the 2HDM under consideration on the different parameters when
scanning over 10000 parameter space points 
 for $M_h=115$ GeV and $M_{H}=500$ GeV.  (Note that $M_A$ and $M_{H^\pm}$ are
free parameters.)}
\label{fig:dsig_dpar_nocut_500}
\end{figure}

\begin{figure}
\center
\epsfig{file=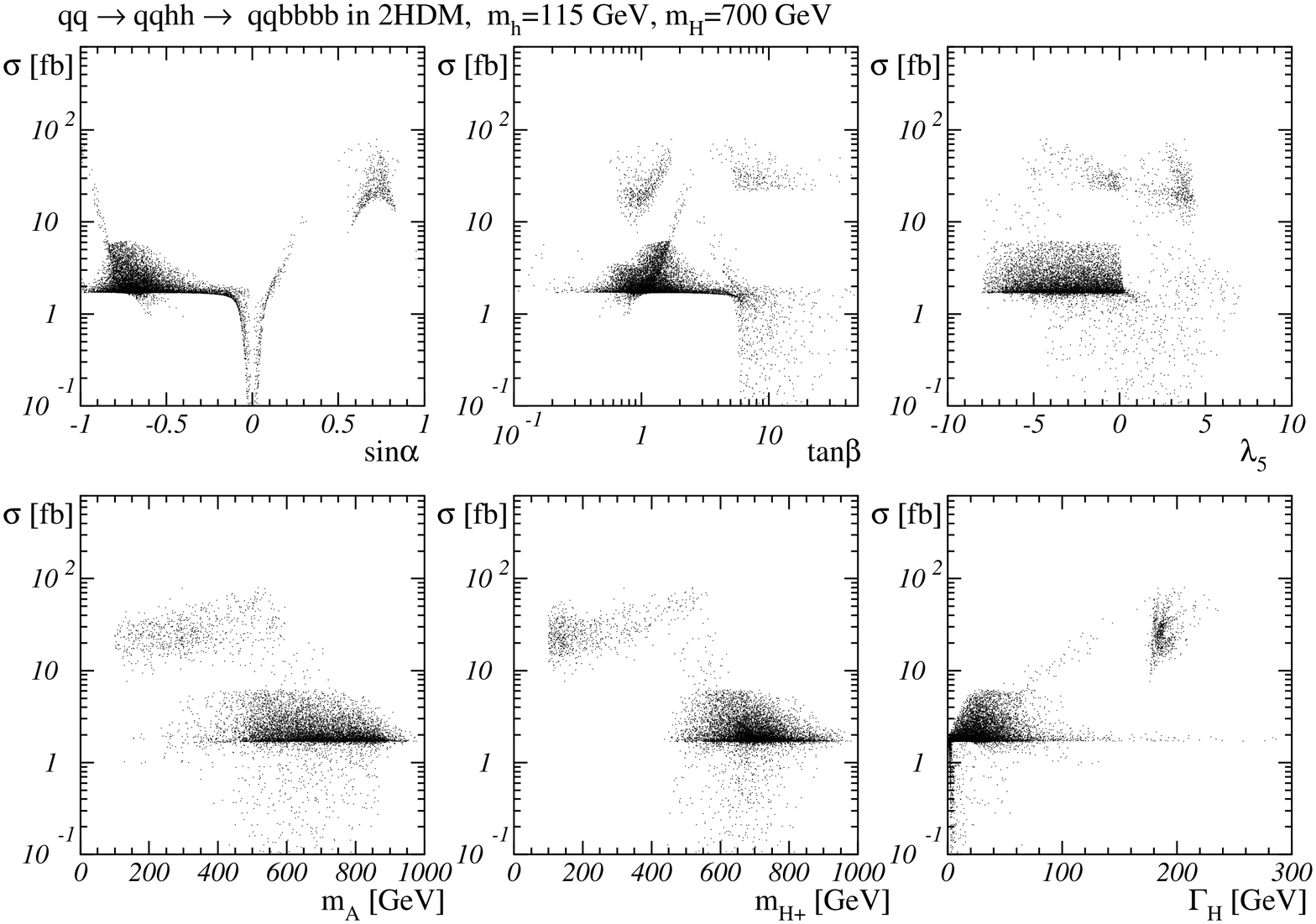,  height=12cm}
\caption{ The dependence of the  inclusive cross-section 
$qq^{(')}\to qq^{(')}hh \to
qq^{(')}b\bar bb\bar b$   in the 2HDM under consideration on the different parameters when
scanning over 10000 parameter space points 
 for $M_h=115$ GeV and $M_{H}=700$ GeV.  (Note that $M_A$ and $M_{H^\pm}$ are
free parameters.)}
\label{fig:dsig_dpar_nocut_700}
\end{figure}

As already alluded to, the parameter space of the general CP-conserving
Type II 2HDM we are considering is quite large as it depends
on seven unknown parameters. In order to get a feel for the dependence of
the signal cross-section for the process 
$qq^{(')}\to qq^{(')}hh \to
qq^{(')}b\bar bb\bar b$  we therefore present in
Figs.~\ref{fig:dsig_dpar_nocut_300} through 
\ref{fig:dsig_dpar_nocut_700} the results of our three
selected scenarios, wherein
 we scan the allowed parameter space over 10000 randomly 
chosen points. 
(Note that similarly to the MSSM case we have included the 
BR($hh \to b\bar bb\bar b$) but not any 
$4b$-jet tagging efficiency.)

Comparing with the cross-sections in the MSSM the main differences are 
due to the following:
\begin{itemize}
\item
the triple Higgs couplings\footnote{We use the same definitions of these
couplings as in \cite{THDM}.} $\lambda_{Hhh}$ and  $\lambda_{hhh}$ are not 
related to the gauge couplings; 
\item
the different parameters can vary independently of each other.
\end{itemize}
Conversely, the kinematic factors in the two models
will be the same for a given set of
masses and widths of the different Higgs bosons. Therefore, in those cases, 
many features of the signal, such as the differential distributions, 
will be similar to those of the MSSM even though the
normalisation can be completely different. 
In fact, comparing Fig.~\ref{fig:scan_qqhh_MSSM_MaxMix} with 
\ref{fig:dsig_dpar_nocut_300} through 
\ref{fig:dsig_dpar_nocut_700} we see that in the more general 2HDM the
cross-sections can be more than two orders of magnitude larger than in the
MSSM thus rendering a
much larger potential for a detectable signal 
(as it will be discussed below). To
be more quantitative on this we give in Tab.~\ref{2hdm-cross-sections} the
maximal inclusive cross-sections obtained in the scans for  
$M_h = 115$~GeV and $M_H = 300$, $500$ and $700$~GeV.

\begin{table}
\begin{center}
\begin{tabular}{|c|l|l|l|l|}\hline
$M_{H}$ (GeV) &  \multicolumn{4}{c|}{$\sigma(qq^{(')}\to qq^{(')}hh) $ [fb] with different cuts} 
\\  \hline
    & inclusive  & primary   & optimal  &  optimal, $H \to hh $ 
 \\ \hline \hline
300 & 1453       & 71.9   & 31.2  & 25.8  
 \\ \hline
500 & 396        & 25.3   & 11.4  &  7.7
 \\ \hline
700 & 80         &  7.1   &  3.3  &  2.0
 \\ \hline
\end{tabular}
\caption{\label{2hdm-cross-sections} The maximal cross-sections
in the 2HDM under consideration 
for $M_h = 115$~GeV, and $M_H = 300$, $500$ and $700$~GeV, respectively, with
the following different cuts: inclusive,  
with primary cuts in
eqs.~(\ref{precuts1})--(\ref{precuts2}), 
and with optimised cuts of eq.~(\ref{eq:optcuts}) in the latter case also when
only considering the $H \to hh$ resonant contribution.}
\end{center}
\end{table}

In order to study the potential signal in more detail we first of all apply 
the same primary cuts as in the case of the MSSM, those listed in 
eqs.~(\ref{precuts1})--(\ref{precuts2}). The resulting cross-sections are
given Tab.~\ref{2hdm-cross-sections}. Comparing with the cross-section
without the primary cuts we see that the reduction is substantial, but even so
the signal cross-section can still be more than two orders of magnitude larger
than in the MSSM scenario considered in subsection~\ref{subsubsec:MSSM} and
it is comparable to the background (see Tab.~\ref{tab:lambda},
specifically for low $\tan\beta$, where the $M_h$ values in the two models
are very similar).

\subsection{Signal-to-Background Differential Analysis}
\label{subsec:S-to-B}

In this section, we will continue the discussion of our numerical 
analyses limitedly to the Type II 2HDM considered so far.
In order to enhance the statistical significance $S/\sqrt{B}$ we 
studied several differential distributions for signals and background 
with the event selection of eqs.~(\ref{precuts1})--(\ref{precuts2}), with 
the aim of introducing optimised cuts, allowing at the same time to keep 
the signal event numbers at a reasonable level. To begin with,
for simplicity, we have limited ourselves to use the contribution from the 
$H \to hh$ resonance to the signal for $M_{H}=300$ GeV in a 
scenario where the cross-section is close to maximal, with 
$\cos(\beta-\alpha)=1$,  $\lambda_{Hhh}=1000$ GeV
and $\Gamma_H=30$ GeV, when
comparing with the background.

\begin{figure}[!t]
\epsfig{file=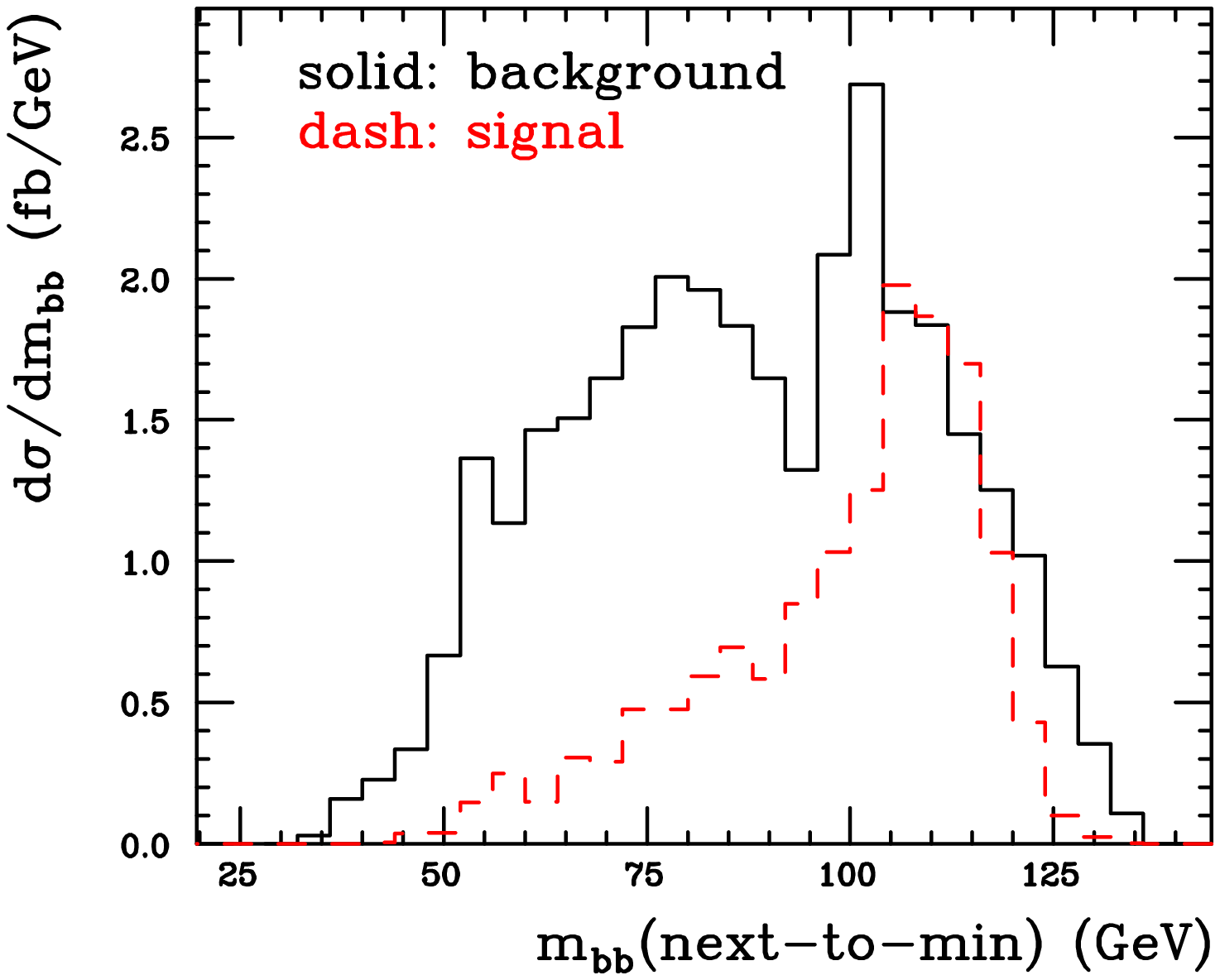,height=6cm,angle=0}
~\hskip0.7truecm\epsfig{file=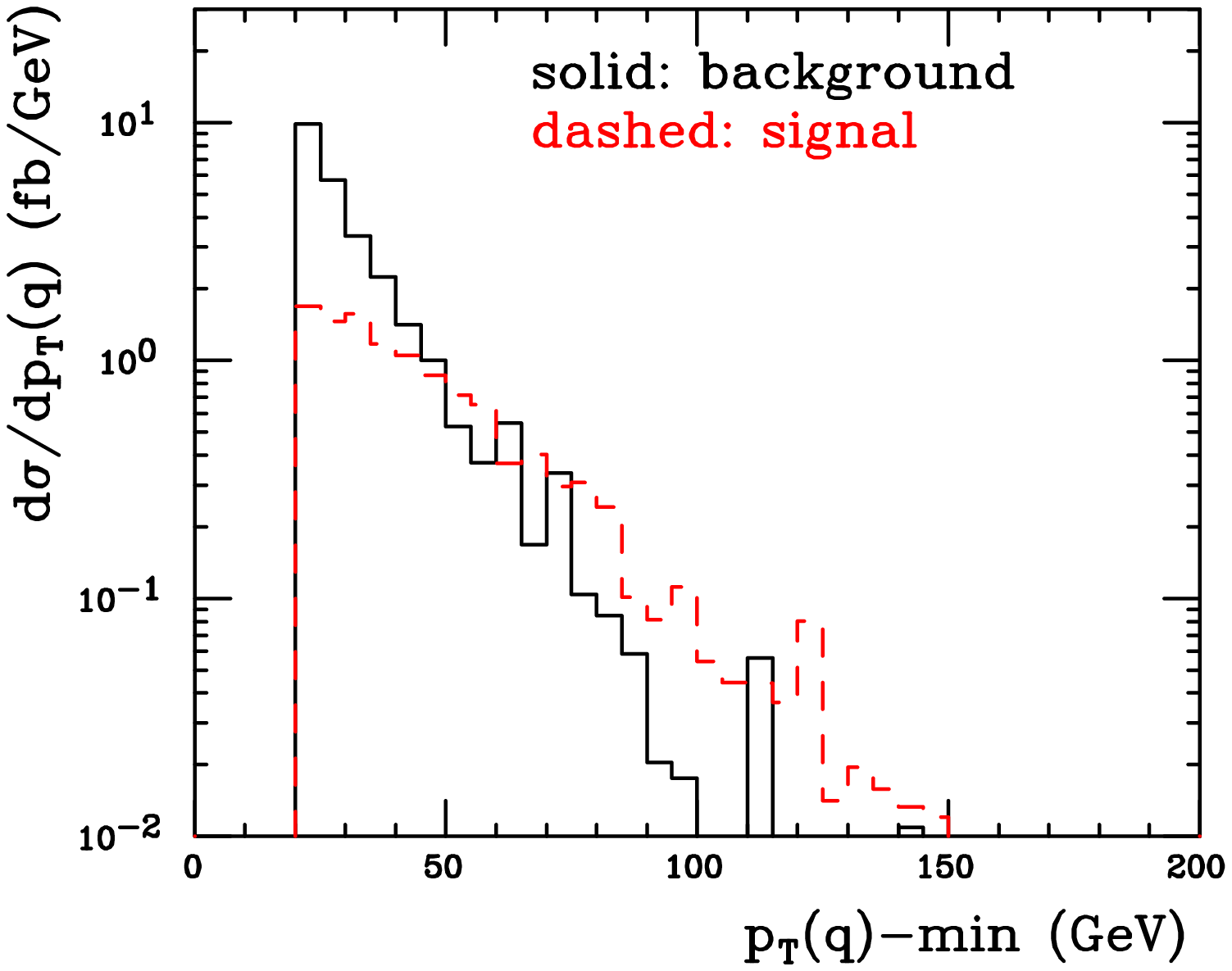,height=6cm,angle=0}
\vspace*{-0.3cm}
\caption{\label{fig:cutsdist}The distribution of 
the next-to-minimum $b \bar b$ invariant 
mass (left) and of the minimum tagging jet transverse momentum (right) 
for the signal (cross section for
$qq^{(')}\to qq^{(')}hh \to
qq^{(')}b\bar bb\bar b$  in a close to best-case scenario for $M_{H}=300$ GeV) 
and the background. The basic cuts of eqs.~(\ref{precuts1})--(\ref{precuts2}) 
are imposed.
} 
\end{figure}

The most sensitive distributions, able to discriminate between the signal
and background, turn out to be the minimum transverse momentum 
of the forward/backward jets and the next-to-minimum invariant mass
of the $b\bar b$ pairs, 
which we show 
in Fig.~\ref{fig:cutsdist}. (Although to a some more limited extent, also the
minimum  $b\bar b$ invariant mass is useful.)  
Before selecting a specific set up, 
we performed also a systematic analysis of the significance 
for different combination of cuts ($30$~GeV $\leq$ $m_{bb}^{\rm min}$ 
$\leq$ $m_{bb}^{\rm next-to-min}$ $\leq 100$~GeV, $20$~GeV $\leq$ 
$p_{\rm T}^{\rm fwd}$ $\leq$ 60~GeV). The best optimised cuts, 
on top of the basic ones of eqs.~(\ref{precuts1})--(\ref{precuts2}),
that we found are\footnote{Note that the efficiency of these is
rather insensitive to
the actual Higgs mass values, so that we have used the same set for any 
choice of the latter.}:
\begin{equation}\label{eq:optcuts}
 p_T^{\rm{fwd/bwd}} >  40~{\rm GeV},\qquad
  m_{bb}^{\rm min} > 40~{\rm GeV},\qquad
 m_{bb}^{\rm next-to-min} > 80~{\rm GeV}.
\end{equation}                                                           

\begin{figure}[!t]
\center
\epsfig{file=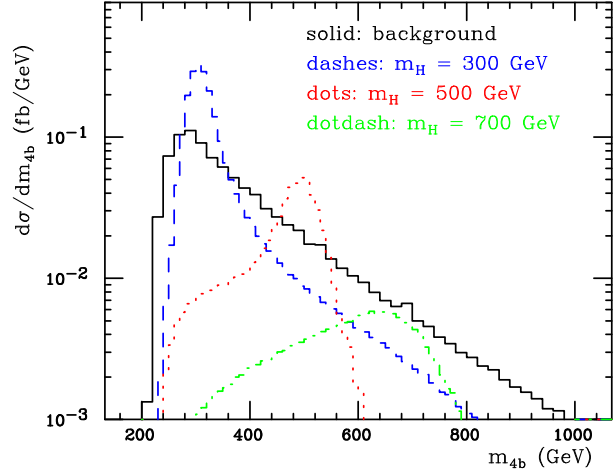,width=8cm,angle=0}
\vspace*{-0.3cm}
\caption{The differential cross-section $d\sigma(qq^{(')}\to qq^{(')}hh \to
qq^{(')}b\bar bb\bar b)/dm_{4b}$ in the best case
scenarios for $M_{H}$ = 300, 500 and 700 GeV obtained when scanning over the
available parameter space restricting the width $\Gamma_H$ to be less 
than 30, 50 and 200 GeV, respectively. 
When calculating the signal distributions the actual
widths have been assumed to be
$\Gamma_H$ = 30, 50 and 200 GeV, respectively.
}
\label{fig:m4bdistr}
\end{figure}

We  show in Fig.~\ref{fig:m4bdistr} the $4b$ invariant mass distribution 
for three signals ($M_H=300$, $500$ and $700$~GeV with the widths
$\Gamma_H$ = 30, 50 and 200 GeV, respectively) and the background 
after the optimised cuts of eq.~(\ref{eq:optcuts}) have also been imposed. 
For each of the three signals shown in the figure we have 
used the parameter space point which gives the maximal signal cross-section
from the resonant $H \to hh$ contribution
when restricting the width $\Gamma_H$ to be less 
than 30, 50 and 200 GeV, respectively.
In this context we note that there are two effects
which mainly determine the width of the signal distribution.
On the one hand,
the smearing of momenta we use gives a contribution to the measurable 
width of about 30 GeV. On the other hand, one of course has 
the intrinsic width of the $H$.

Taking suitable mass windows around the peaks for the different 
Higgs mass values
illustrated in Fig.~\ref{fig:m4bdistr}, 
we obtain the maximal signal cross-sections, event numbers and 
statistical significances quoted in Tab.~\ref{significances}. In order to
calculate the signal cross-sections in the respective windows for different
parameter space points, taking the actual width of the $H$ into account,
we rescaled the contribution from the $H \to hh$ resonance with a factor
$c_{M_{H}} \equiv \left(\arctan\left[2(M_{H}-m_L)/\Gamma_i\right] + 
                       \arctan\left[2(m_U-M_{H})/\Gamma_i\right] \right)$ 
where $m_L$ and $m_U$ 
are the lower and upper limits of the signal window, $\Gamma_i$ is the width
of the signal distribution in parameter space point $i$ estimated from
$\Gamma_i=\sqrt{\Gamma_{H_i}^2+\Gamma_{\rm 4b}^2}$ with 
$\Gamma_{\rm 4b}= 30 $ GeV being the width of the
$m_{4b}$-distributions from finite detector resolution.
(Notice then that $c_{M_{H}}$ 
is a normalisation determined from scenarios with 
$\Gamma_{H_i}$ = 30, 50 and 
200 GeV for the different $H$ masses.) 
Thus we approximate the cross-section in the  $m_{4b}$ window as 
$\sigma_{\rm peak} = c_{M_{H}} \sigma_{H\to hh}$.
As a further requirement we
also imposed that at least 50\% of the signal
cross-section after the optimal cuts comes from the $H\to hh$ resonance
such that the would-be-signal would not be
obscured by other non-resonant
contributions. 

\begin{figure}
\center
\epsfig{file=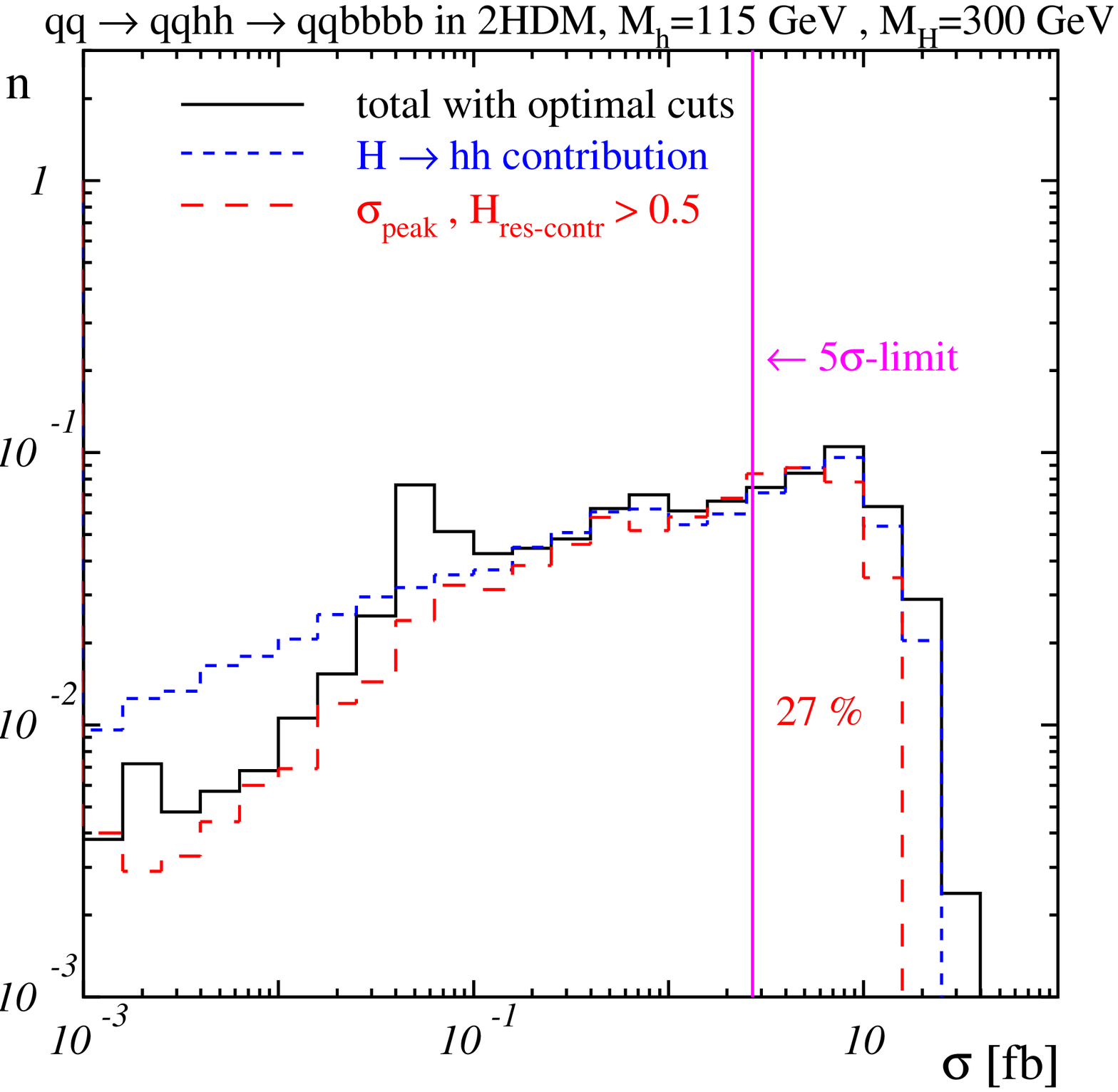,  height=8.4cm}
\epsfig{file=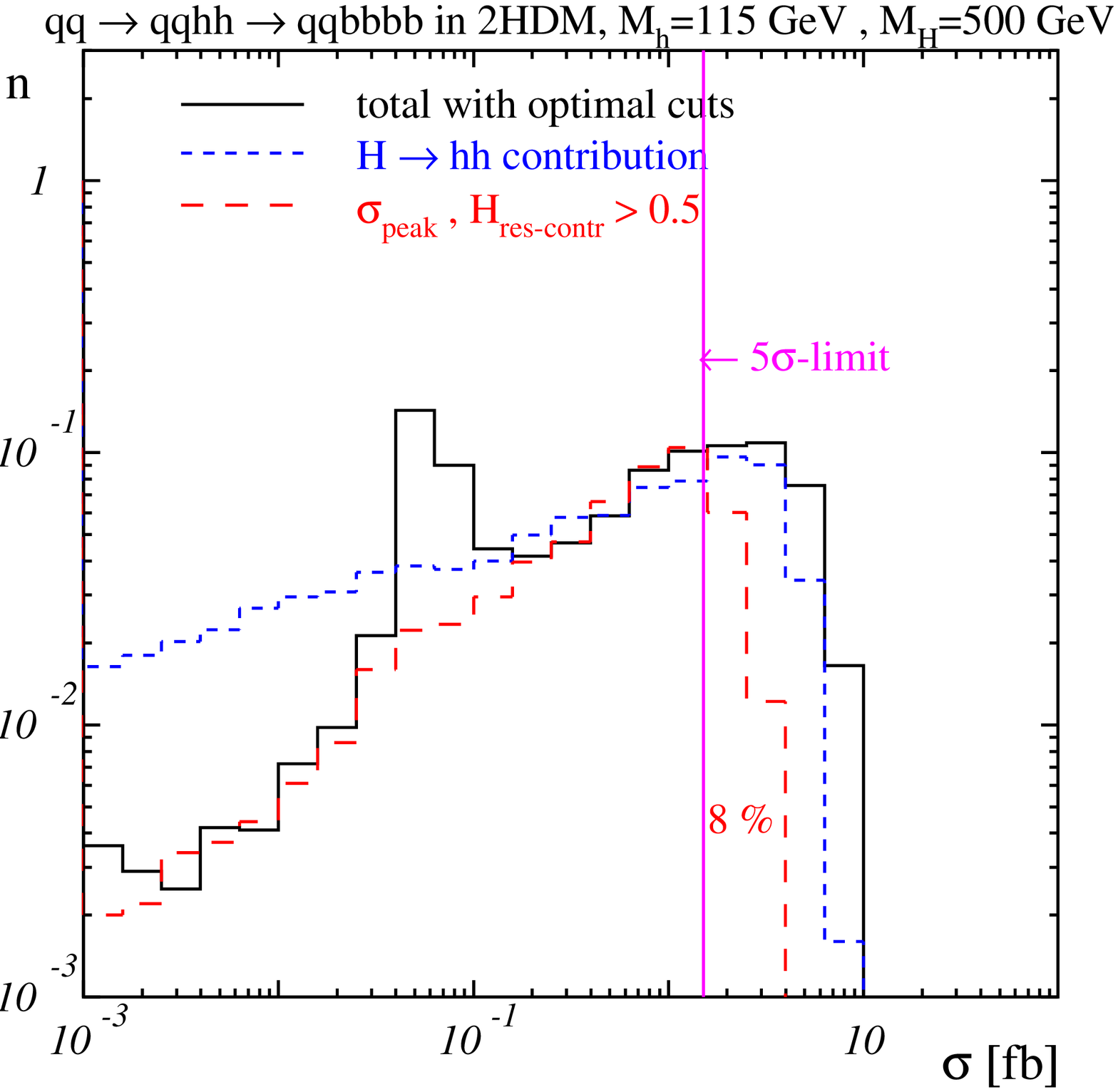,  height=8.4cm}
\epsfig{file=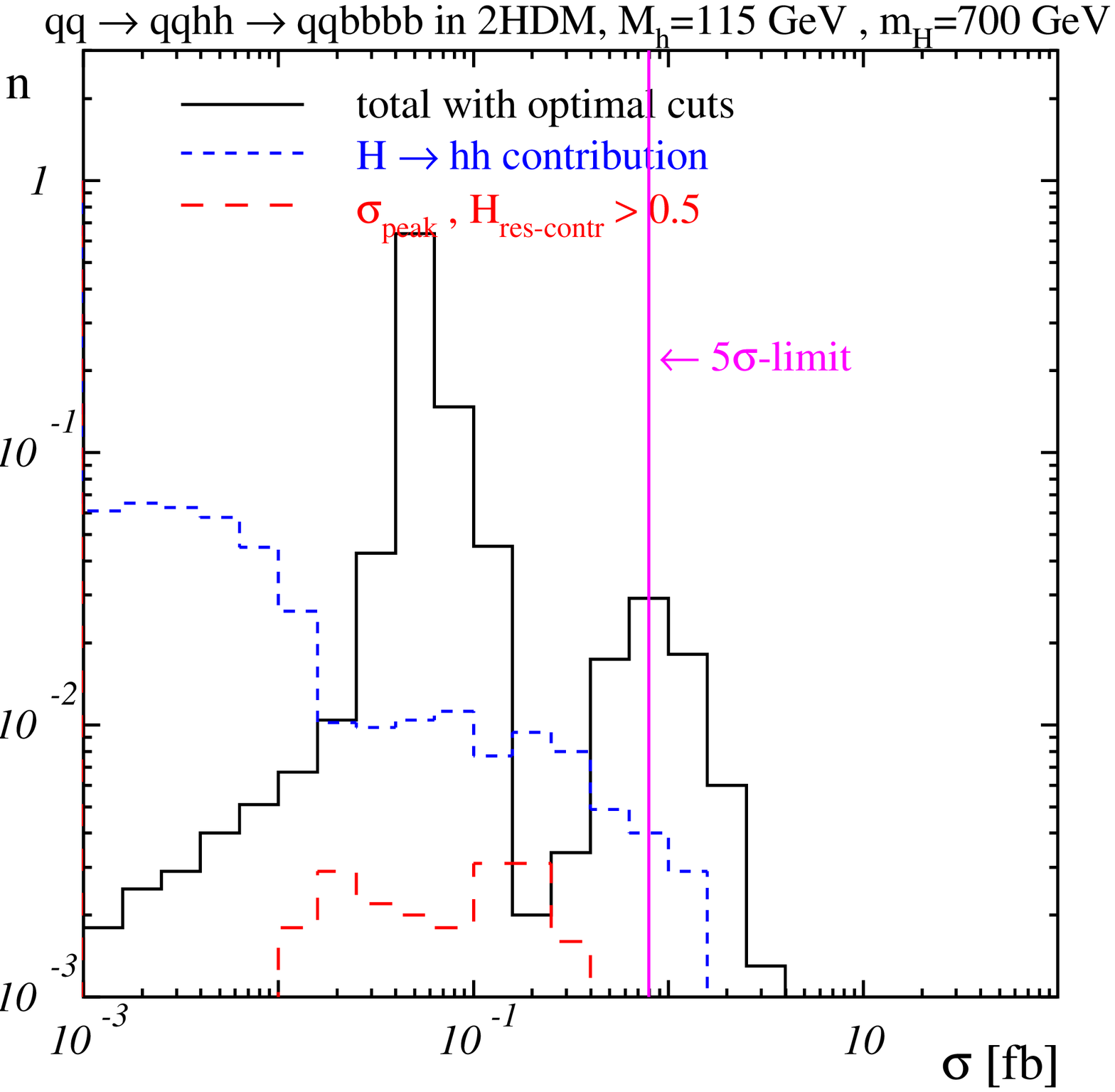,  height=8.4cm}
\caption{ Distributions of the resulting cross-sections 
$qq^{(')}\to qq^{(')}hh \to
qq^{(')}b\bar bb\bar b$   in the 2HDM under consideration 
using the  optimal cuts obtained in a scan 
over 10000 parameter space points (the area is normalised to 1 for the 
cross-section with optimal cuts) 
for three different sets of Higgs boson masses as indicated in 
the respective plots. The solid line shows the results
with optimal cuts, the dashed line shows the resonant contribution from the $H\to
hh$ processes and the long dashed line shows the resonant contribution in the
respective signal windows requiring that at least 50\% of the
cross-section comes from the $H\to hh$ resonance. 
The vertical line corresponds to the $5\sigma$-limit at LHC assuming 
300~fb$^{-1}$ and the
integral of the curves to the right of it gives the
percentage of parameter space points where the resonant cross-section is 
larger than this.
}
\label{fig:dlogn_dsig}
\end{figure}

The distributions of the signal cross-sections obtained in this
way are given in Fig.~\ref{fig:dlogn_dsig}. For illustration, the
$5\sigma$ limits at LHC, assuming an integrated luminosity of 300 fb$^{-1}$, 
$\sigma_{\rm peak}>$ 2.7 (2.3-3.3)\footnote{The ranges given within parenthesis 
in this paragraph have been obtained by varying the 
factorisation/renormalisation scale for the background by a factor of 
two around the default value, 
which makes the corresponding cross-section decrease by 
30 \% or increase by 50 \% respectively. The reason for this is that,
being essentially a six-jet cross-section, the background rate is proportional to
$\alpha_s^6$ and it is therefore quite sensitive to the renormalisation scale. 
We also note that our default scale ($p_T^2=\sum_{1}^{n}p_{Tj}^2/n$) has 
conservatively been chosen to be small so, if anything, our estimate of the 
final signal-to-background rates should be regarded as conservative. 
In a real experiment one should of course attempt to use the
sidebands for background normalisation.}, 1.5 (1.3-1.8) and 0.8 (0.7-1.0) fb, 
for $M_H=300$, $500$ and $700$~GeV respectively, 
are also illustrated and the fractions of parameter space 
points which gives cross-sections larger then this are  
27 (24-30), 8 (5-12) and 0\%, respectively. 
The corresponding numbers for the SLHC with 3000 fb$^{-1}$ are
43 (41-45), 31 (28-33) and 2 (0-2)\%, respectively. Thus even at the SLHC 
we find no scope of observing a $M_H=700$~GeV resonance in the channel under
investigation.

\begin{figure}
\center
\epsfig{file=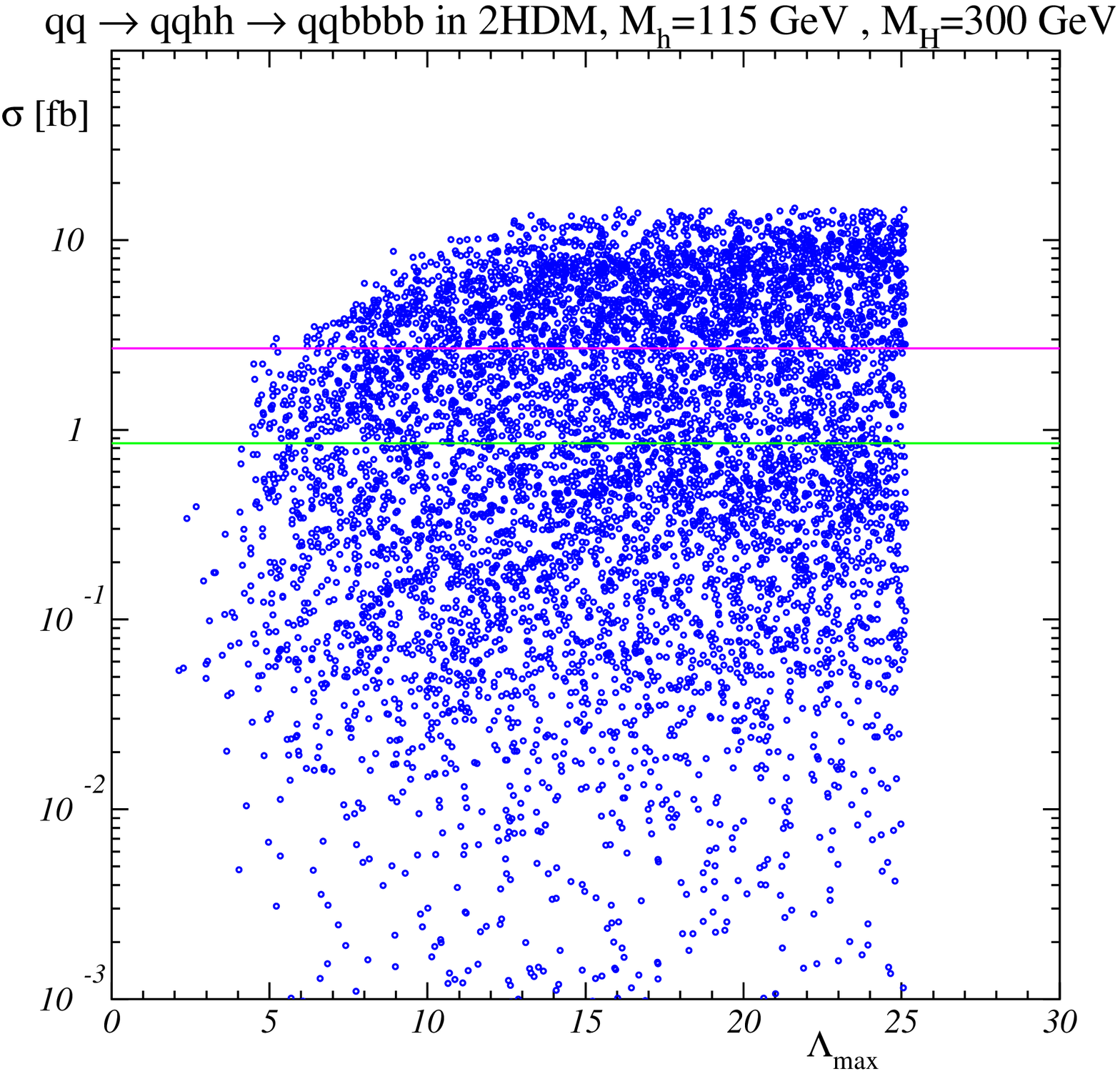,  height=8.4cm}
\epsfig{file=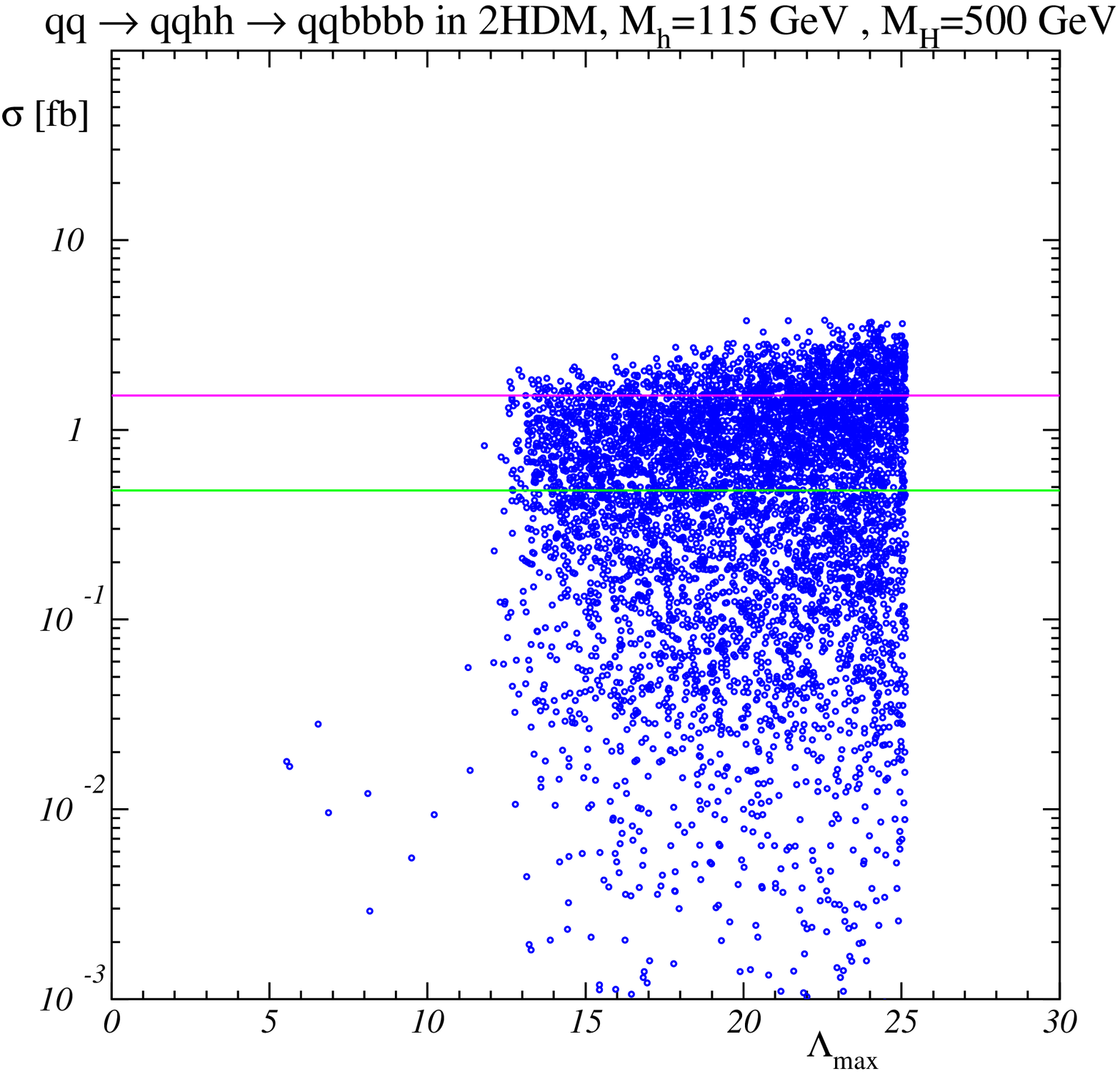,  height=8.4cm}
\epsfig{file=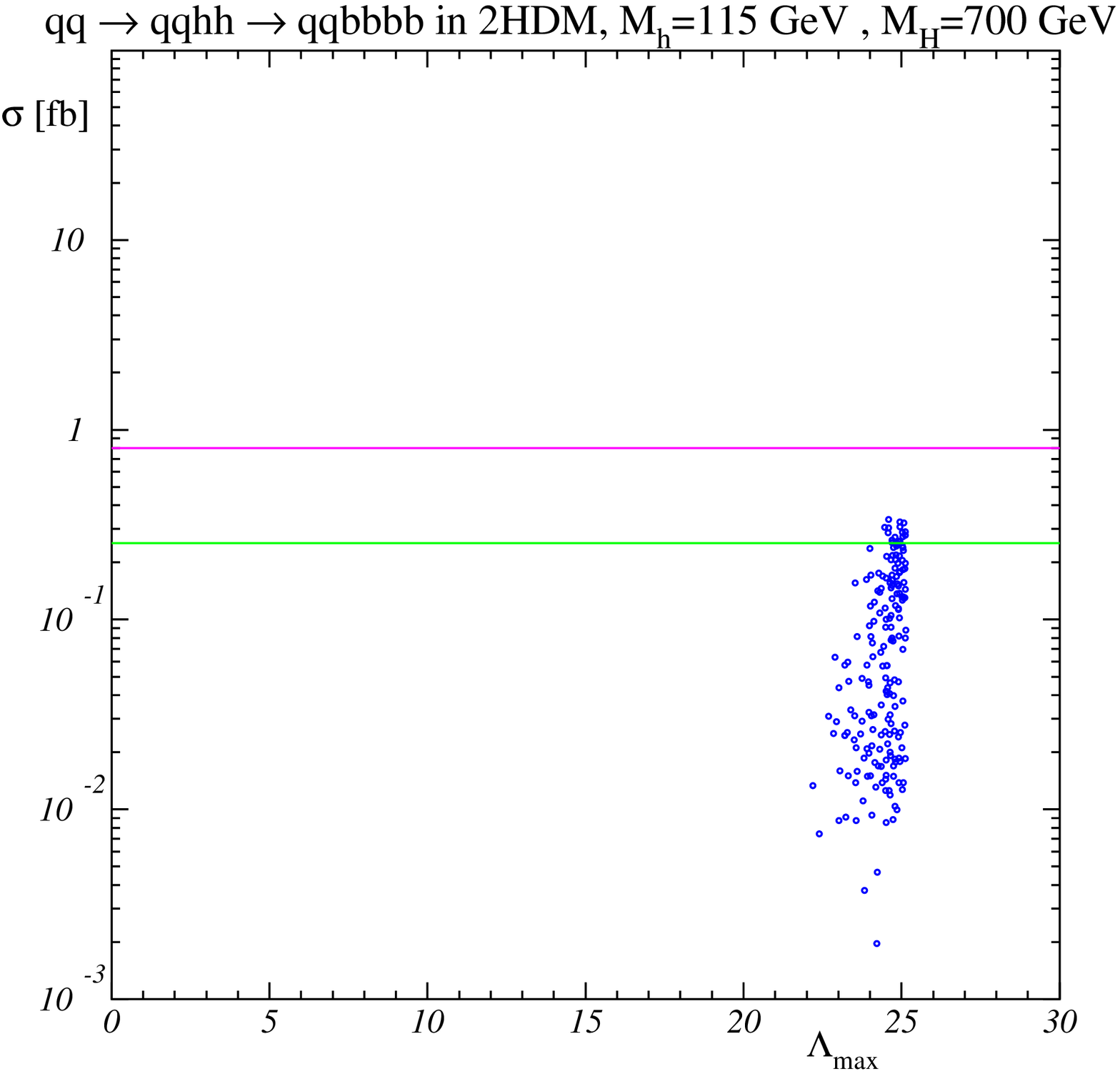,  height=8.4cm}
\caption{ The distributions in resulting signal cross-sections 
$qq^{(')}\to qq^{(')}hh \to
qq^{(')}b\bar bb\bar b$   in the 2HDM under consideration 
using the  optimal cuts obtained in a scan 
over 10000 parameter space points as a function of the maximal eigenvalue
$\Lambda_{\max}$ of the scattering matrix 
for three different sets of Higgs boson masses as indicated in 
the respective plots.  The upper (lower) horizontal line corresponds to 
the $5\sigma$-limit at (S)LHC assuming 300~fb$^{-1}$ (3000~fb$^{-1}$).
}
\label{fig:dsig_dlam}
\end{figure}

Finally we have also investigated the effects
of restricting the allowed parameter space from tree-level unitarity by 
putting harder constraints on the maximal eigenvalue of the scattering 
matrices, $\Lambda_{\max}$. For this purpose, Fig.~\ref{fig:dsig_dlam}
shows the signal cross-sections obtained in the scan as a function of  
$\Lambda_{\max}$. From the figure it is clear that the results (at least for 
$M_H=300$ and 500 GeV) are not sensitive
to the precise value used for applying the unitarity constraint. On the other
hand, applying a much harder constraint of the order $\Lambda_{\max} \lsim 4 (12)$ 
(instead of  $\Lambda_{\max} < 8 \pi $) essentially leads to that the 
sensitivity for detection at the LHC is more or less washed out for 
$M_H=300$ (500) GeV. The same also holds at the SLHC assuming an integrated 
luminosity of 3000 fb$^{-1}$.

\begin{table}
\begin{center}
\begin{tabular}{|l|l|l|l|l|l|}\hline
 $m_{4b} $ window & $B$ events &  $\sigma_{\rm peak}^{\max}$ [fb]&  $S$ events& $S/\sqrt{B}$@LHC & $S/\sqrt{B}$@SLHC
\\  \hline
280 -- 340 (GeV) & 102  & 15.1   &  283    & 28  &  89
 \\ \hline
460 -- 540 (GeV) & 30   &  3.8   &   71    & 13  &  41
 \\ \hline
660 -- 740 (GeV) & 8    &  0.35  &   6.6   & 2.3  &  7.4
 \\ \hline
\end{tabular}
\caption{\label{significances} Number of events and significances 
for $M_h = 115$~GeV and $M_H = 300$, $500$, $700$~GeV in the respective best
case scenarios, for a
$4b$-tagging efficiency of (50\%)$^4$ and after
the optimised cuts of eq.~(\ref{eq:optcuts}). 
The assumed integrated luminosity at LHC and SLHC are 300~fb$^{-1}$ 
and 3000~fb$^{-1}$, 
respectively. }
\end{center}
\end{table}

\section{Conclusions}
\label{sec:summary}
We would like to conclude our paper by stating that, at both the
LHC and SLHC,  there exists a great potential to extract a $H\to hh\to 4b$
resonance when $M_h$ is constrained
in the vicinity of 115 GeV.
This is a crucial result if one recalls that 
 the detection of a sole Higgs resonance and consequent 
extraction of an $M_h$ value 
may not point unambiguously to the underlying model of EWSB, 
not even in presence of further measurements of the heavier
Higgs masses, $M_H$, $M_A$ and/or $M_{H^\pm}$. 

For example,
the 2HDM considered here may be realised in a configuration
wherein all visible Higgs masses are degenerate with those
of the MSSM. Under these circumstances, we have proved that
\begin{itemize}
\item it is not possible to extract an $H\to hh\to 4b$ resonance
from
vector-boson fusion in the MSSM (not even if $M_H$ is known) 
whilst
\item the opposite case is true in a substantial fraction of the parameter space
of our 2HDM (even if $M_H$ is not known), thereby
enabling one to possibly measure the triple-Higgs coupling 
$\lambda_{Hhh}$. 
\end{itemize}
The latter is a Lagrangian term, which is
different between these two models even when their patterns of Higgs 
masses and couplings to SM objects are the same, 
that would give a unique insight into the underlying EWSB mechanism.

To be more specific our results show that in the most favourable scenario with 
$M_H=300$ GeV up to 27 (43) \% of the parameter space would give a $5\sigma$ 
signal at the (S)LHC assuming an integrated luminosity of 300 (3000) fb$^{-1}$ 
when using the standard tree-level unitarity requirement on the $J=0$ partial 
waves, ${\rm Re} (a_0) < 1/2$. These results are not sensitive to the precise 
value used for applying the unitarity constraint, albeit for very strong 
constraints the sensitivity for detecting the signal goes away. In the case of 
$M_H=500$ GeV the fraction of parameter space probed is smaller with  
up to 8 (31) \% giving a $5\sigma$ signal, whereas for $M_H=700$ GeV there is
essentially no sensitivity at all.

Despite we lack a full Monte Carlo simulation we believe
to have incorporated the
most critical aspects of the latter so that we do not  expect  
more realistic studies (including parton shower, hadronisation, 
heavy hadron decays and detector effects)
to affect too strongly our conclusions. 

Finally, we are currently pursuing other work along the directions
outlined here, covering the case of lightest (neutral) 
Higgs boson pair production
in the case of Higgs-strahlung and in association with heavy quarks 
\cite{preparation}.

\section*{Acknowledgments} We are all grateful to the (formerly)
CERN Theory Division for hospitality when this work was started.
Several discussions with M.L. Mangano are acknowledged.
SM and FP thank JR for his kind hospitality in Uppsala in September 2006.
FP thanks SM for his stay in Southampton in March and June 2007.
RP acknowledges the financial support of the MIUR under contract
2006020509\_004 and of the RTN European Programme
MRTN-CT-2006-035505 (HEPTOOLS, Tools and Precision Calculations
for Physics Discoveries at Colliders). SM acknowledges the latter too
for partial funding. 
RP's research was partially supported by the ToK Program
"ALGOTOOLS" (MTKD-CT-2004-014319). FP 
thanks the CERN Theory Unit for partial support.

\def\pr#1 #2 #3 { {\rm Phys. Rev.} {\bf #1} (#2) #3}
\def\prd#1 #2 #3{ {\rm Phys. Rev. D} {\bf #1} (#2) #3}
\def\prl#1 #2 #3{ {\rm Phys. Rev. Lett.} {\bf #1} (#2) #3}
\def\plb#1 #2 #3{ {\rm Phys. Lett. B} {\bf #1} (#2) #3}
\def\npb#1 #2 #3{ {\rm Nucl. Phys. B} {\bf #1} (#2) #3}
\def\prp#1 #2 #3{ {\rm Phys. Rep.} {\bf #1} (#2) #3}
\def\zpc#1 #2 #3{ {\rm Z. Phys. C} {\bf #1} (#2) #3}
\def\epjc#1 #2 #3{ {\rm Eur. Phys. J. C} {\bf #1} (#2) #3}
\def\mpl#1 #2 #3{ {\rm Mod. Phys. Lett. A} {\bf #1} (#2) #3}
\def\ijmp#1 #2 #3{{\rm Int. J. Mod. Phys. A} {\bf #1} (#2) #3}
\def\ptp#1 #2 #3{ {\rm Prog. Theor. Phys.} {\bf #1} (#2) #3}
\def\jhep#1 #2 #3{ {\rm JHEP} {\bf #1} (#2) #3}
\def\jphg#1 #2 #3{ {\rm J. Phys. G} {\bf #1} (#2) #3}
\def\cpc#1 #2 #3{ {\rm Comp. Phys. Comm.} {\bf #1} (#2) #3} 

\clearpage

\clearpage

\clearpage

\clearpage

\clearpage

\clearpage

\clearpage

\clearpage

\clearpage

\begin{thebibliography}{99}


\bibitem{guide} J.F.~Gunion, H.E.~Haber, G.L.~Kane and S.~Dawson,
                ``The Higgs Hunter Guide''
                (Addison-Wesley, Reading MA, 1990),
                {Erratum}, 
                  {\tt hep-ph/9302272}.

\bibitem{THDM} J.F. Gunion and H.E. Haber,
Phys. Rev. {\bf D67} (2003) 075019.

\bibitem{Fawzi} F. Boudjema and A.V. Semenov, Phys. Rev. {\bf D66} (2002)
 095007.

\bibitem{Abdel}
  A.~Djouadi,
  arXiv:hep-ph/0503172;
  arXiv:hep-ph/0503173.


\bibitem{HHH} M.~Moretti, S.~Moretti, F.~Piccinini, R.~Pittau and A.D.~Polosa,
JHEP {\bf 02} (2005) 04; {\tt hep-ph/0411039}.


\bibitem{ATLAS} 
ATLAS collaboration, 
`{ATLAS Technical proposal}', 
CERN/LHCC/94-43, LHCC/P2, 1994;
`ATLAS Detector and Physics Performance Technical
Design Report', ATLAS TDR 14, CERN/LHCC 99-14, 1999.


\bibitem{CMS} 
CMS collaboration, 
`{CMS Technical Proposal}',
CERN/LHCC/94-38, LHCC/P1, 1994;
`CMS Physics Technical Design Report, Volume I:
Detector Performance and Software'
CMS TDR 8.1,
CERN/LHCC 2006-001, 2006.   

\bibitem{SLHC} 
F. Gianotti, M.L. Mangano and T. Virdee (conveners),
{\tt hep-ph/0204087}.

\bibitem{VVHH} W.-Y. Keung, Mod. Phys. Lett. {\bf A10} (1987) 765;
        O.J.P. \'Eboli, G.C. Marques, S.F. Novaes and A.A. Natale,
        Phys. Lett. {\bf B197} (1987) 269;
        D.A.~Dicus, K.J.~Kallianpur and S.S.D.~Willenbrock, Phys.\
        Lett.\ {\bf B200} (1988) 187;
        K.J.~Kallianpur, Phys.\
        Lett.\ {\bf B215} (1988) 392;
        A.~Abbasabadi, W.W.~Repko,
        D.A.~Dicus and R.~Vega, Phys.\ Rev.\ {\bf D38} (1988) 2770,
        Phys.\ Lett. {\bf B213} (1988) 386;
        A.~Dobrovolskaya and V.~Novikov, Z.\ Phys.\ {\bf C52} (1991)
        427.

\bibitem{ggHH}
        E.W.N.~Glover and J.J.~van der Bij, Nucl.\ Phys.\ {\bf B309}
        (1988) 282;
D.A. Dicus, C. Kao and S.S.D. Willenbrock, Phys. Lett. {\bf B203} (1988)
457; G.~Jikia, {Nucl.~Phys.} {\bf B412} (1994) 57;
T. Plehn, M. Spira and P.M. Zerwas,
 Nucl. Phys. {\bf B479} (1996) 46, Erratum, ibidem {\bf B531}
(1998) 655.

\bibitem{Remi} R. Lafaye, D.J. Miller, M. Muhlleitner and
 S. Moretti, 
{\tt hep-ph/0002238}.

\bibitem{LesHouches1999}
A. Djouadi, R. Kinnunen, E. Richter-Was, H.U. Martyn {\it et al.},
in 
{\tt hep-ph/0002258}.

\bibitem{standard} E. Richter-Was {et al.}, Int. J. Mod Phys. {\bf
    A13} (1998) 1371; E. Richter-Was and D. Froidevaux, Z. Phys. {\bf
    C76} (1997) 665; J. Dai, J.F. Gunion and R. Vega, Phys. Lett. {\bf
    B371} (1996) 71, {ibidem}, {\bf B378} (1996) 801.


\bibitem{Baur1} U. Baur, T. Plehn and D. Rainwater,
Phys. Rev. {\bf D68} (2003) 033001. 

\bibitem{Remi2} S. Balatenychev et al., in {\tt hep-ph/0203056}.

\bibitem{Binoth:2006ym}
  T.~Binoth, S.~Karg, N.~Kauer and R.~Ruckl,
  Phys.\ Rev.\  {\bf D74} (2006) 113008.

\bibitem{Djouadi:1999rca}
  A.~Djouadi, W.~Kilian, M.~Muhlleitner and P.~M.~Zerwas,
  Eur.\ Phys.\ J.\  C {\bf 10} (1999) 45.
  

\bibitem{hepdata} \url{http://durpdg.dur.ac.uk/hepdata/pdf.html}.

\bibitem{ALPGEN} M.L. Mangano, M. Moretti, F. Piccinini, R. Pittau
and A.D. Polosa,
JHEP {\bf 07} (2003) 001.

\bibitem{HELAS}
H.~Murayama, I.~Watanabe and K.~Hagiwara,
                       KEK Report 91--11, January 1992.

\bibitem{VEGAS} G.P. Lepage, J. Comp. Phys. {\bf 27} (1978) 192,
preprint CLNS-80/447, March 1980.

\bibitem{Metropolis} H. Kharraziha and S. Moretti,
{\rm Comp. Phys. Comm.} {\bf 127} ({2000})
 {242}, Erratum, ibidem {\bf 134} (2001) 136.

\bibitem{MaxMix} M. Carena, P.H. Chankowski, S. Pokorski and C.E.M. Wagner,
 Phys. Lett. {\bf B441} (1998) 205.

\bibitem{HDECAY} A.~Djouadi, J.~Kalinowski and M.~Spira,
{\rm Comp. Phys. Comm.}  {\bf 108} ({1998}) {56}.

\bibitem{Akeroyd:2000wc}
A.G.~Akeroyd, A.~Arhrib and E.M.~Naimi,
Phys.\ Lett. {\bf B490} (2000) 119.

\bibitem{Ginzburg:2005dt}
  I.~F.~Ginzburg and I.~P.~Ivanov,
  Phys.\ Rev.\  D {\bf 72} (2005) 115010.

\bibitem{charged_Higgs_pairs} 
S.~Moretti and J.~Rathsman,
Eur.\ Phys.\ J.\ {\bf C33} (2004) 41.

\bibitem{BRs} S. Moretti and W.J. Stirling, Phys. Lett. {\bf B347} 
({1995}) {291}, Erratum, {ibidem} {\bf B366} (1996) 451.

\bibitem{HiggsLEP2}
See:
\url{http://lephiggs.web.cern.ch/LEPHIGGS/papers/}.



\bibitem{preparation} M.~Moretti, S.~Moretti, F.~Piccinini, R.~Pittau and 
J.~Rathsman, in preparation.

\end{thebibliography}
\end{document}